\documentclass[%
reprint,
superscriptaddress,
amsmath,amssymb,
aps,longbibliography,
prx,
]{revtex4-2}

\usepackage{graphicx}
\usepackage{verbatim}
\usepackage[version=4]{mhchem}
\usepackage{xcolor}
\usepackage{physics}
\usepackage{tabularx}
\usepackage{booktabs}
\usepackage[normalem]{ulem}
\usepackage{pdfpages}
\usepackage{pgffor}

\usepackage{array}
\usepackage{siunitx}
\usepackage{booktabs}
\usepackage{multirow}
\usepackage{array}
\usepackage{tabularx}
\usepackage{float}

\newcommand{\editor}[2]{%
  \expandafter\newcommand\csname #1note\endcsname[1]{%
  \textcolor{#2}{(\textbf{#1note:} \textsc{##1})}}%
  \expandafter\newcommand\csname #1\endcsname[1]{%
  \ifhighlight\textcolor{#2}{##1} \else ##1\fi}%
  \expandafter\newcommand\csname #1add\endcsname[1]{%
  \textcolor{#2}{##1}}%
  \expandafter\newcommand\csname #1cancel\endcsname[1]{%
  \ifhighlight\textcolor{#2}{\sout{##1}}\fi}%
  \expandafter\newcommand\csname #1change\endcsname[2]{%
  \ifhighlight\textcolor{#2}{\sout{##1} ##2}\else ##2\fi}%
  \newenvironment{#1text}{\ifhighlight\color{#2}\fi}{\color{black}}
}
\definecolor{tangerine}{rgb}{0.944,0.522,0}
\definecolor{verde}{rgb}{0.,0.6,0}

\editor{resub}{cyan}
\editor{MC}{purple}
\editor{JA}{blue}
\editor{CM}{tangerine}
\editor{PP}{verde}
\editor{AM}{magenta}
\editor{PL}{red}

\makeatletter
\AtBeginDocument{\let\LS@rot\@undefined}
\makeatother

\begin{document}

\title{High-quality, high-information datasets\\for universal atomistic machine learning} 

\author{Cesare~Malosso}
\affiliation{Laboratory of Computational Science and Modeling, Institut des Mat\'eriaux, \'Ecole Polytechnique F\'ed\'erale de Lausanne, 1015 Lausanne, Switzerland}

\author{Filippo~Bigi}
\affiliation{Laboratory of Computational Science and Modeling, Institut des Mat\'eriaux, \'Ecole Polytechnique F\'ed\'erale de Lausanne, 1015 Lausanne, Switzerland}

\author{Paolo~Pegolo}
\affiliation{Laboratory of Computational Science and Modeling, Institut des Mat\'eriaux, \'Ecole Polytechnique F\'ed\'erale de Lausanne, 1015 Lausanne, Switzerland}

\author{Joseph~W.~Abbott}
\affiliation{Laboratory of Computational Science and Modeling, Institut des Mat\'eriaux, \'Ecole Polytechnique F\'ed\'erale de Lausanne, 1015 Lausanne, Switzerland}

\author{Philip~Loche}
\affiliation{Laboratory of Computational Science and Modeling, Institut des Mat\'eriaux, \'Ecole Polytechnique F\'ed\'erale de Lausanne, 1015 Lausanne, Switzerland}

\author{Mariana~Rossi}
\affiliation{MPI for the Structure and Dynamics of Matter, 22761 Hamburg, Germany}
\affiliation{Yusuf Hamied Department of Chemistry, University of Cambridge, CB2 1EW Cambridge, UK}

\author{Tiago~J.~Goncalves}
\affiliation{BASF SE, Group Research Carl-Bosch-Straße 38, D-67056, Ludwigshafen, Germany}

\author{Sandip~De}
\affiliation{BASF SE, Group Research Carl-Bosch-Straße 38, D-67056, Ludwigshafen, Germany}

\author{Michele~Ceriotti}
\email{michele.ceriotti@epfl.ch}
\affiliation{Laboratory of Computational Science and Modeling, Institut des Mat\'eriaux, \'Ecole Polytechnique F\'ed\'erale de Lausanne, 1015 Lausanne, Switzerland}

\author{Arslan~Mazitov}
\affiliation{Laboratory of Computational Science and Modeling, Institut des Mat\'eriaux, \'Ecole Polytechnique F\'ed\'erale de Lausanne, 1015 Lausanne, Switzerland}

\newcommand{\CSM}[1]{{\color{red}#1}}

\date{\today}

\begin{abstract}
The quality, consistency, and information content of training data is often what determines the practical value of machine-learning models for atomistic simulations. 
Yet, many widely used electronic-structure databases are assembled having materials screening as primary goal rather than robust force-field learning, are limited in their scope to a specific class of chemical compounds, and/or employ inconsistent DFT functionals and settings. 
Here we introduce MAD-1.6, a highly curated dataset designed explicitly for training atomistic models that are broadly applicable across the periodic table at high levels of theory.
MAD-1.6 extends the MAD dataset with targeted enrichment strategies that improve the coverage of chemical space to 102 elements while keeping the total number of configurations compact. 
All structures are computed with a single, standardized all-electron DFT workflow using the r$^2$SCAN meta-GGA functional and consistent convergence settings, ensuring uniformity across chemically heterogeneous systems. 
The dataset encompasses molecules, clusters, bulk crystals, surfaces, low-dimensional structures, molecular clusters and adsorbed molecules, and its quality and consistency are further enhanced by outlier removal using uncertainty quantification.
We demonstrate the high accuracy that can be achieved with the proposed dataset by training PET-MAD-1.6, a generally applicable r$^2$SCAN interatomic potential that covers 102 elements in the periodic table and achieves exceptional levels of benchmark accuracy and stability in challenging simulation protocols.

\end{abstract}

\maketitle

\section{Introduction}
Atomistic simulations increasingly rely on machine-learned surrogate models to bridge the gap between first-principles accuracy and the length- and time-scales needed for effective modeling~\cite{Behler2016,Thiemann2024,Jacobs2025}. While modern architectures can represent complex potential-energy surfaces, their reliability in practice is often limited by the training data. The coverage of chemical space, the diversity of local environments, and the presence of challenging off-equilibrium configurations in a dataset can determine accuracy in molecular dynamics, structure optimization, and in the description of rare events, of reactions and of interfacial processes~\cite{mazitov2025pet,mad-dataset}.

Training broadly transferable interatomic models raises two recurring data challenges. First, many available reference collections are dominated by near-equilibrium structures, which result in models that are not sufficiently constrained in the distorted, high-force, and close-contact regimes that control stability in challenging modeling scenarios, especially at high temperatures. 
Second, datasets assembled over long periods or from multiple sources often mix electronic-structure settings in subtle ways (e.g., numerical thresholds, treatment of magnetism, or other minor protocol choices), resulting in small, but at times significant, inconsistencies.

Recent learning-oriented datasets address these limitations by explicitly targeting both coverage and consistency~\cite{wbm, alexandria, omat, omol, matpes}. 
However, they fail in achieving a fully universal character in different ways, either because they are limited to a class of chemical compounds (e.g., only extended systems or only molecules), or because they employ electronic-structure settings that are inconsistent between different classes of compounds. 
Furthermore, large datasets can often contain redundant atomic environments, diluting their information content.

In Ref.~\citenum{mad-dataset} some of us introduced MAD, a dataset based on the principle of Massive Atomic Diversity -- incorporating different classes of highly-distorted structures, and using consistent and convergent electronic-structure settings, based on density-functional theory (DFT) at the PBEsol~\cite{perd+08prl} exchange-correlation level. 
To further build on these ideas, we now  introduce MAD-1.6, an extension of MAD (from now on referred to as MAD-1) designed to increase chemical completeness and interaction coverage while retaining compactness and reference consistency. 
All configurations are computed with a single standardized all-electron DFT workflow using the r$^2$SCAN meta-GGA functional~\cite{Sun2015,r2scan}, which has shown excellent transferability among different chemistries \cite{sun+nchem2016,kotha+acs2023,Liu2024}. 
MAD-1.6 expands element coverage up to 102 elements, including every isotope with a half-life above a day, and enriches under-represented portions of chemical space through targeted additions. The dataset is built iteratively, making early versions of the data and of corresponding trained machine learning interatomic potentials available for testing and application by others, and using them internally to inform the generation of new data sections.
We characterize the final dataset and demonstrate its utility by training a general-purpose machine-learned interatomic potential, PET-MAD-1.6, observing exquisite robustness and transferability across diverse materials and molecular environments.

\begin{figure*}
\centering
\includegraphics[width=0.7\linewidth]{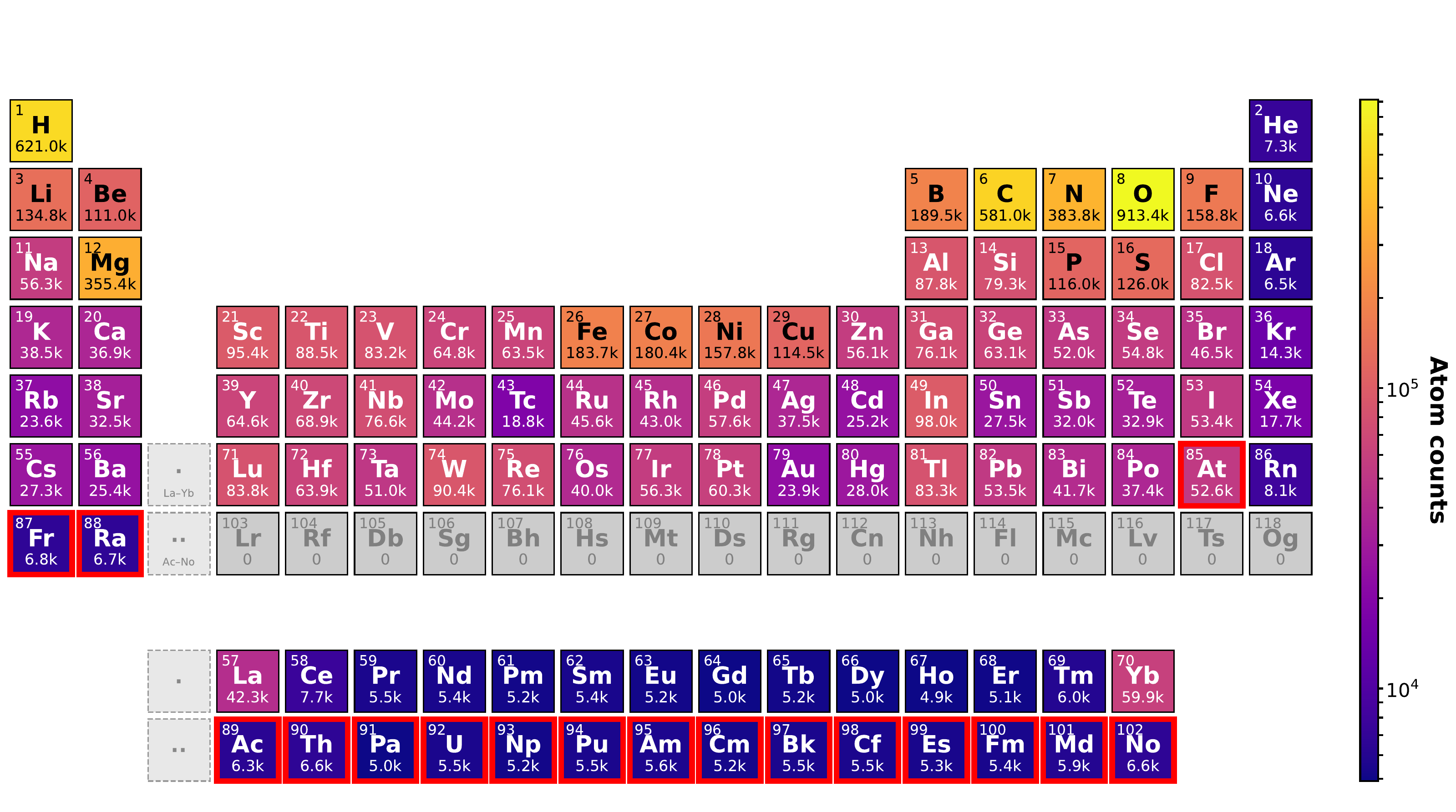}
    \caption{Periodic table indicating the statistical representation of the elements present in the MAD-1.6 dataset. Each element tile contains the atomic number, element symbol, and elemental frequency in MAD-1.6, by which it is also colored. Elements in gray are not present in the dataset, and those bordered in red have been newly introduce to MAD-1.6 compared to the original MAD-1 dataset.}\label{fig:periodic_table}
\end{figure*}

\section{Dataset construction}

\subsection{Composition}

The MAD-1.6 dataset contains 362646 atomic structures spanning 102 chemical elements. It is constructed as an extension of the MAD-1 dataset, augmented with structures specifically designed to enrich the coverage and predictive accuracy across the entire periodic table. The element counts of the MAD-1.6 dataset are shown in Fig.~\ref{fig:periodic_table}. 
The dataset is organized into 16 subsets (summarized in Table~\ref{tab:subsets}) that span a wide range of chemical compositions and structural motifs, including molecules, clusters, bulk crystals, low-dimensional materials, and surfaces. The structures inherited from MAD-1 (8 subsets) provide a broad baseline of chemically and structurally diverse configurations, including 85 elements, while the newly introduced structures (8 new subsets) systematically target under-represented elements, coordination environments, and bonding regimes.
We briefly describe the individual subsets comprising MAD-1.6, outlining their makeup, and how they contribute to extend chemical and configurational coverage. Since we built the dataset iteratively, using MLIPs trained on the intermediate versions to inform the final makeup, we list separately the composition of these intermediate versions, which remain available online for reference and reproducibility (\url{https://github.com/lab-cosmo/upet}).

\begin{table*}[tb]
\centering
\footnotesize
\begin{tabular}{lllrr}
\toprule
 & Subset name & Description & \# structures & \# atoms \\
\midrule

\multirow{8}{*}{MAD-1 dataset}
& MC3D & Bulk crystals from the Materials Cloud 3D (MC3D) database & 33 343 & 728 155 \\
& MC3D-rattled & Rattled configurations derived from MC3D & 29 214 & 569 254 \\
& MC3D-random & Atomic species randomized over 85 elements in MC3D structures  & 2 075 & 17 052 \\
& MC3D-surface & Surface slabs generated from MC3D crystals & 4 928 & 181 674 \\
& MC3D-cluster & Nanoclusters cut from MC3D and MC3D-rattled structures & 8 651 & 42 867 \\
& MC2D & Two-dimensional crystals from the Materials Cloud 2D database & 2 502 & 40 950 \\
& SHIFTML-molcrys & Curated molecular crystals from the SHIFTML dataset & 8 572 & 851 596 \\
& SHIFTML-molfrags & Neutral molecular fragments from SHIFTML & 3 242 & 72 130 \\

\midrule

\multirow{6}{*}{New in MAD-1.5}
& monomers & Isolated single-atoms covering all 102 elements & 102 & 102 \\
& dimers & Isolated two-atom configurations sampled from 102 elements & 31 325 & 62 650 \\
& trimers & Isolated three-atom configurations sampled from 102 elements & 2 306 & 6 918 \\
& MC3D-extended & Extended MC3D crystals covering 93 elements with $Z<100$ & 11 562 & 442 266 \\
& MC3D-random-extended & Atomic species randomized over 102 elements in MC3D structures & 5 397 & 54 472 \\
& binary-random & Binary FCC and BCC random decorations over 102 elements & 73 584 & 201 958 \\

\midrule
\multirow{1}{*}{New in MAD-1.6} 
& trimers-extended & Extension of the isolated three-atom configurations & 92 278 & 276 834 \\
& MAD-cat & Reaction pathways of adsorbates on curated metal surfaces & 53 565 & 3 901 110 \\

\midrule
{\bf Total MAD-1.6} &  &  & {\bf 362 646} & {\bf 7 449 988} \\

\bottomrule

\end{tabular}
\caption{Composition of the MAD-1.6 dataset. Original MAD-1 dataset is augmented with isolated atoms, dimers, trimers, extended and randomized crystals, and randomized binary structures to systematically expand chemical and configurational diversity up to 102 elements. The total number of structures and atoms across all subsets is reported in the last row.}
\label{tab:subsets}
\end{table*}
For a comprehensive description of the structures inherited from MAD-1, we refer the reader to the original publication~\cite{mad-dataset}. Beyond explicit inclusion of \emph{monomers}, new structures introduced in MAD-1.6 span seven distinct subsets, each designed to sample different coordination environments and bonding regimes. Analogous to the original \emph{Materials Cloud 3D Database} (MC3D)~\cite{MC3D} dataset, the \emph{MC3D-extended} subset contains 11562 periodic structures, but including a broader variety of elements, covering previously missing lanthanides and actinides, as well as additional under-represented main-group and transition metals.
The \emph{MC3D-random-extended} subset addresses the limited representation of heavy elements in MAD-1. It applies the same randomization strategy as \emph{MC3D-random}~\cite{mad-dataset}, but with the added constraint that each structure includes at least one of the previously missing or under-represented elements. 
This approach ensures that the interactions involving these elements are adequately sampled while keeping the total number of configurations minimal and the diversity maximal. The \emph{binary-random} subset contains binary substitutional orderings generated by decorating supercells of BCC and FCC parent lattices with all pairs of chemical species drawn from the 102 elements. 
We restrict sampling to supercells containing up to three parent-lattice sites and, for each supercell, enumerate all distinct site-occupation patterns compatible with a binary system~\cite{Hart2008, Kolli2020}.
Including this set ensures broad chemical coverage and provides a strong prior on interactions across the full set of elements (including chemically atypical pairings) while avoiding the combinatorial growth associated with larger cells and more complex orderings. 
Cell volumes are chosen such that the average volume per atom is equal to the average van der Waals volume of the elements in the system.

The \emph{dimers} subset explicitly samples two-body interactions across the entire periodic table by systematically generating all unique pairs of 102 elements. For each element pair, interatomic distances are sampled along ten linearly spaced points between a minimum distance of 0.125 times the sum of the van der Waals radii and a maximum distance equal to the sum of the van der Waals radii. This sampling scheme captures the repulsive regime at short ranges, the equilibrium bonding region, and the onset of the long-range tail, providing comprehensive potential-energy curves for all possible elemental pairings. 
We exclude larger distances because of the known self-interaction and static-correlation issues of non-spin-polarized DFT for open-shell systems in this regime~\cite{cohe+08science, bryenton2023delocalization}. 
The inclusion of these isolated dimers contributes to a reliable prediction of short-range repulsive interactions and low-coordination bonding environments that are under-represented in extended structures, particularly for element combinations that rarely occur in experimentally observed compounds.

The \emph{trimers} and \emph{trimers-extended} subsets extend this coverage to three-body interactions. The former comprises configurations obtained by drawing three elements at random from the 102 available and sampling positions such that every interatomic distance lies between $0.25$ and $1.0$ times the sum of the covalent radii of the pair, with a fixed radius of 1.6~\AA\ assumed for elements with $Z \geq 97$. Random selection of the elements, however, leaves much of composition space unvisited. The \emph{trimers-extended} subset therefore enumerates systematically all unordered element combinations, allowing repetition, with one configuration per combination. Here atom $A$ is placed at the origin while $B$ and $C$ are placed along directions drawn uniformly on the unit sphere, at distances sampled from a ten-point uniform grid spanning $0.25$ to $1.0$ times the sum of the van der Waals radii of the corresponding pair; a configuration is accepted only if the resulting $B$--$C$ separation exceeds $0.25$ times the sum of the van der Waals radii of $B$ and $C$, leaving the third distance bounded from below but free to extend beyond contact. Together, the dimers, trimers and trimers-extended subsets provide explicit low-body-order references that anchor the model's representation of fundamental interatomic interactions~\cite{chong+26jcp}, complementing the higher-body-order environments present in bulk, surface, and molecular configurations.

To strengthen the description of reactive events and adsorbate--surface interactions relevant to heterogeneous catalysis --- configurations largely absent from the bulk, surface, and molecular subsets --- MAD-1.6 introduces the \emph{MAD-cat} subset, which samples minimum-energy reaction pathways of small C-, N-, and O-containing intermediates on metal surfaces. 
Starting from a library of adsorbed C$_1$/C$_2$ and N$_1$/N$_2$ species, along with oxygenated and O/H molecules (e.g.\ CO, CO$_2$, CHO, CH$_2$O, CH$_3$OH, CH$_4$, CH$_3$CHO, CH$_3$CH$_2$OH, NH, NH$_2$, NH$_3$, N$_2$H$_2$, N$_2$, OH, H$_2$O, H$_2$, O$_2$), we enumerate their bond-scission reactions (dehydrogenation and C--O/C--C/N--O/N--N/O--O cleavage) on the close-packed Cu(111) terrace and the stepped Cu(211) facet, taken as representatives of coordinatively saturated and under-coordinated (step/edge) active sites, respectively. Every reaction is processed with the PET-MAD-1 potential from Ref. \cite{mazi+25ncomm}: each transition state is located with a ARPESS search~\cite{arpess}, characterized by vibrational analysis, and relaxed along its imaginary mode toward both adjacent minima in an intrinsic-reaction-coordinate (IRC)-like procedure. 
From each pathway we retain the converged saddle point together with up to eight configurations sampled along the two descending branches --- the path minima, the valley--ridge inflection points, and farthest-point-sampled intermediate images --- discarding any structure with a maximum atomic force above 15~eV/\AA{}.

To extend this reactive sampling across the periodic table, each adsorbate and
transition-state motif is abstracted into a marker-anchored molecular
fragment using \emph{AutoAdsorbate}~\cite{autoadsorbate} and re-deposited onto a diverse ensemble of catalytic surfaces derived from the \emph{MC3D-surface} pool. Slabs unsuited as catalytic models (thickness below 8~\AA{}, fragmented or cluster-like slabs, and pure non-metal or actinide compositions) are removed, and 492 monometallic and binary surfaces are selected using the farthest-point sampling (FPS) algorithm applied over smooth overlap atomic positions (SOAP) descriptors ~\cite{de+16pccp, jage+18npjcm, bart+13prb, will+18pccp}.
Each slab is expanded to a lateral size $\gtrsim 9$~\AA{}, its lower half is frozen, and a small random displacement of 0.05~\AA{} is applied to enrich the force
distribution. 
Fragment re-deposition over symmetry-distinct adsorption sites, followed by geometric sanity checks (adsorbate bound to the surface, no sub-surface penetration), yields $\approx$ 70{,}000 candidate configurations, which are then recomputed with the standardized all-electron r$^2$SCAN workflow used throughout MAD-1.6 and resulting in $\approx$ 54{,}000 converged calculations.

Overall, MAD-1.6 combines the diversity of MAD-1 with targeted structural enrichment and the improvement of the level of reference DFT theory, resulting in a dataset that offers broad chemical coverage across the full periodic table, compact in size, but rich in chemical information.

\subsection{Electronic-structure details}

All calculations were performed with FHI-aims \cite{AIMS}, an all-electron, numeric atom-centered orbital (NAO) electronic-structure code~\cite{FHI-aims2025}. Version 250806 was used for the majority of the dataset. A subset of structures was recomputed with version 260622 to correct a bug in version 250806 that affected the stress tensor for meta-GGA functionals; for these structures, all target properties were taken from the version-260622 calculations. Since the bug affected only the stress tensor, the remaining target properties are identical between the two versions.
The input parameters are standardized across the dataset, enforcing consistency on the results. As the exchange–correlation functional, we opt for the meta-GGA r$^2$SCAN \cite{Sun2015,r2scan}. This functional improves accuracy significantly in comparison to the GGAs used in most of the available DFT datasets (including the original MAD-1 dataset), while keeping the computational cost affordable. In particular, r$^2$SCAN was shown to vastly improve formation enthalpies of a wide range of solids~\cite{kotha+acs2023} and, like SCAN, improve the description of hydrogen-bonds~\cite{Chen2017,sun+nchem2016} with respect to GGAs.

To achieve robust convergence, we employ an 8~\AA{}$^{-1}$ k-point density grid (for periodic systems) and a Gaussian electronic smearing of 0.05 eV. We enforce convergence thresholds for the self-consistent solution of the Kohn–Sham equations of $10^{-6}$ eV for the energy, $10^{-4}$ eV/\AA~for the forces, and $10^{-5}$ $e\cdot a_0^{-3}$ for the electron density.
FHI-aims ``tight'' basis sets and numerical settings are used for all elements. Species defaults (2020 version) provided by FHI-aims are used throughout, with a slight modification: For a subset of lanthanide and actinide elements (Pr–Yb, Pu–No, excluding Ce), we removed the confined 5d/6d function because, although physically motivated, they were found to hinder SCF convergence in non-spin-polarized calculations. This choice has limited impact on the accuracy of the resulting energies and forces. 
Each calculation outputs total energies, atomic forces, and stress tensors for periodic systems. 
We emphasize that using non-spin-polarized calculations neglects important physical effects, and affects the accuracy and numerical stability of calculations, e.g. for $f$-block elements and for dissociation curves. As for the MAD-1 dataset, we maintain that in the absence of a universally applicable DFT framework that converges reliably to the ground state of arbitrary materials, internal consistency has to be prioritized when building a dataset that strives to be truly universal.

\subsection{Outlier detection}
\label{sec:dataset-cleaning}

Achieving consistent DFT targets at the r$^2$SCAN meta-GGA level across a large dataset of 102 elements is highly non-trivial, primarily due to degeneracies in self-consistent field (SCF) calculations and the sensitivity of this functional to integration grids~\cite{r2scan}. These issues increase the risk of obtaining DFT ground states that do not converge or that converge to a local minimum, leading to substantial differences in calculated targets in repeated runs. 

\begin{table}[tb]
    \centering
    \begin{tabular*}{\columnwidth}{@{\extracolsep{\fill}}lrr}
    \toprule
    Subset & Heuristic filter & LLPR filter \\
    \midrule
    MC3D & 33 357 & 33 343 \\
    MC3D-rattled & 29 228 & 29 214 \\
    MC3D-random & 2 117 & 2 075 \\
    MC3D-surface & 4 958 & 4 928 \\
    MC3D-cluster & 8 981 & 8 651 \\
    MC2D & 2 529 & 2 502 \\
    SHIFTML-molcrys & 8 574 & 8 572 \\
    SHIFTML-molfrags & 3242 & --- \\
    \midrule
    monomers & 102 & --- \\
    dimers & 34 215 & 31 325 \\
    trimers & 2 481 & 2 306 \\
    MC3D-extended & 11 679 & 11 562 \\
    MC3D-random-ext. & 5 574 & 5 397 \\
    binary-random & 78 010 & 73 584 \\
    trimers-extended & 105 208 &  92 278 \\
    MAD-cat & 53 567 & 53 565\\
    \midrule
    Total & 383 822 & 362 646 \\
    \bottomrule
    \end{tabular*}
    \caption{MAD-1.6 subsets counts after successive outlier detection steps. Heuristic-based cleaning removed all structures with force magnitudes $>$ 100 eV/\AA. LLPR-uncertainty-based cleaning involved training and calibrating a LLPR model on the heuristically-cleaned dataset, and further eliminates the structures for which the actual absolute error in energy prediction was 3 times higher than the predicted energy uncertainty. A dash `---' in the LLPR column indicates no change due to LLPR-based cleaning.}
    \label{tab:dataset-cleaning}
\end{table}

In order to mitigate this effect during model training, we performed a two-step cleaning procedure of the initial dataset of successfully converged calculations. First, we applied a heuristic-based cleaning, similar to that we used for MAD-1~\cite{mazi+25sd}. We filtered out all structures with force magnitudes greater than 100 eV/\AA \ to prevent overfitting on energies dominated by short-range repulsive contributions. 
Next, we used an uncertainty-based algorithm to eliminate potentially inconsistent structures with low predicted uncertainty and high actual error, following a procedure similar to that used in Ref.~\citenum{paru+18ncomm} to remove poorly converged chemical shielding calculations.
To accomplish this, we trained the preliminary ML potential on the heuristically cleaned dataset from the previous step, and we calibrated a LLPR energy uncertainty estimate on the same training data (using a S-size model, see Section~\ref{sub:model-arch} and Section~\ref{sub:model-uq} for details).
We then filtered out all structures for which the predicted energy uncertainty was 3 times lower than the actual error. This procedure was performed in two independent iterations. During the first iteration, both LLPR model training and data cleaning were performed on the MAD-1.5 dataset. The second iteration involved only the new subsets added to the MAD-1.6 dataset, namely \emph{trimers-extended}, and \emph{MAD-cat}. A visualization of both steps of the LLPR cleaning process is shown in Figure \ref{fig:llpr-cleaning}. The overall statistics on the subsets counts before and after the cleaning is shown in Table \ref{tab:dataset-cleaning}. 
We also distribute the set of structures that have been discarded by the LLPR cleaning procedure, both to compare with other dataset normalization protocols, and as they are likely to constitute challenging cases to stress-test the convergence of DFT implementations.

\section{Models and benchmarks}

\subsection{Model architecture and training}
\label{sub:model-arch}

To demonstrate the accuracy that can be achieved using a diverse, internally consistent dataset, we train a machine-learning interatomic potential based on the the \emph{Point Edge Transformer} (PET) \cite{pozdnyakov2023smooth, bigi2026pushing}. 
PET is a rotationally unconstrained, transformer-based graph neural network (GNN) designed to predict the properties of atomistic systems with high accuracy and computational efficiency.
The model predicts energies as a sum of atomic contributions. Conservative forces and stresses are computed using automatic differentiation as derivatives of the total energy prediction with respect to atomic positions and strain tensors, respectively. 
We use separate heads to predict non-conservative forces and stresses, which can be used to accelerate simulations~\cite{bigi+25icml}.
We refer the reader to Ref.~\citenum{bigi2026pushing}, that discusses the specific architecture we use here, including an adaptive cutoff approach that allows to achieve a balanced description of systems of widely different density. 
In fact, the PET-MAD-1.6 models presented in this work are fine-tuned from those trained in Ref.~\citenum{bigi2026pushing} on the OMat24 dataset~\cite{barroso2024open}, and they therefore match the XS, S, and M sizes of the models presented there with minor changes discussed below.

  \begin{table*}
      \centering
      \small
      \begin{tabular}{lccc}
      \toprule
      Model & XS & S & M \\
      \midrule
      Trained from & PET-OMat-XS & PET-OMat-S & PET-OMat-M \\
      Parameter count & 4.5M & 25.9M & 108.6M \\
      Node features & 512 & 1024 & 1536 \\
      Edge features & 128 & 256 & 384 \\
      GNN layers & 2 & 3 & 3 \\
      Attention layers & 1 & 1 & 2 \\
      Cutoff radius (\AA) & 7.5 & 8.0 & 8.5 \\
      Cutoff width (\AA) & 1.0 & 1.0 & 1.0\\
      Adaptive \# neighbors & 16 & 24 & 32 \\
      \midrule
      Max learning rate & 2e-4 & 2e-4 & 1e-4 \\
      Batch size  (\#  atoms) & 1024 & 512 & 512 \\
      Number of epochs & 200 & 200 & 200 \\
      Weight decay & 1e-5 & 1e-4 & 1e-5 \\
      E loss weight & 10.0 & 10.0 & 10.0 \\
      F loss weight &  1.0  & 1.0 & 1.0 \\
      S loss weight &  1.0  & 1.0 & 1.0 \\
      NC F loss weight &  0.1  & 0.1 & 0.1 \\
      NC S loss weight &  0.1  & 0.1 & 0.1 \\
      \bottomrule
      \end{tabular}
      \caption{Model and training hyperparameters for the models evaluated in this work. Three model sizes of the PET-MAD-1.6 (XS,
  S and M) were trained from the corresponding OMat checkpoints from the Ref.~\citenum{bigi2026pushing}, where the the adaptive number of neighbors parameter along with the cutoff function width were overwritten with the values in the current table.  The number of node
  features corresponds to four times that of edge features for all models. The batch size is reported per GPU and corresponds to a total number of atoms targeted by a batch sampler. In all cases, a linear learning rate warm-up of 1\% of
  the total number of epochs was performed, followed by cosine learning rate annealing, which starts from the maximum value and
  decays to zero at the end of training.  Both r$^2$SCAN and PBE targets were trained with the same weights for energies (E),
  forces (F), stresses (S), non-conservative forces (NC F) and non-conservative stresses (NC S) in the loss function. }
      \label{tab:hyperparameters}
  \end{table*}

All the models were trained on a cleaned version of the MAD-1.6 dataset (cleaning details are provided in Sec. \ref{sec:dataset-cleaning}), targeting the atomization energy of each system - i.e. the quantity left after subtracting stoichiometric monomer energies from the total energy, atomic forces and, where applicable, stresses. The dataset was split into training, validation, and test subsets comprising roughly 80\%, 10\%, and 10\% of the total number of cleaned structures, respectively, using a stratified split method. Since dimers and trimers define the fundamental two- and three-body interactions and, together with monomers, the zero-density limit of the potential, inserting these structures in the training set ensures physically correct short- and long-range behavior and prevents the model from compensating errors through spurious many-body correlations~\cite{chong+26jcp}. 
Like monomers, these low-order clusters are not included to benchmark generalization; rather, they are incorporated as physical constraints in the training split, increasing its relative size to 87\%. 
We observed that including a separate head targeting a lower level of theory improves the force accuracy of the fine-tuned model by approximately 25\% -- possibly by further reducing the impact of some residual inconsistencies of the harder-to-converge r$^2$SCAN functional and/or the mismatch in level of theory with the pre-trained model. 
For this reason, we also used a subset of the MAD-1.6 dataset, that we obtained using the PBE functional \cite{Perdew1996} during early phases of dataset construction, split in accordance with the cross-validation subsets mentioned above. These data are targeted with separate, PBE-specific heads and equivalent loss weightings as described in Table~\ref{tab:hyperparameters}.
Given that this dataset has a lower level of theory, and it is not as carefully curated as MAD-1.6, we discard the PBE heads from the models after training. 
The training runs were performed using the \texttt{metatrain} package~\cite{bigi2025metatensor} based on PyTorch~\cite{pasz+19nips} using 16, 16, and 32 NVIDIA GH200 GPUs, and took approximately 7, 13, and 11 hours for the XS, S, and M sizes, respectively. A batch sampler that targeted the total number of atoms in the batch instead of a total number of structures was used to enable smoother training behavior and avoid memory spikes caused by imbalances in the dataset's structure size. All the hyperparameters used during model selection and training are listed in Table \ref{tab:hyperparameters}. Weight optimization was performed using the AdamW optimizer \cite{adamw} with a weight decay of $10^{-5}$ combined with the cosine annealing learning rate scheduler \cite{cosineannealing} and a linear learning rate warm-up for the first 10\% of the total number of epochs.
Prior to training, targets are normalized in the following ways: 1) a further dataset-dependent chemical composition-based contribution is subtracted from the atomization energies, and 2) targets are scaled by dividing each by their standard deviation across the training set. The loss function was based on a weighted sum of the root-mean-squared errors for each predicted target. Non-conservative heads for direct forces and stresses predictions were trained along with their conservative counterparts for both r$^2$SCAN and PBE heads with a reduced weight in the loss function, in order to enable accelerated inference~\cite{bigi+25icml}. The loss weightings of each target are provided in Table~\ref{tab:hyperparameters}.

\subsection{Uncertainty quantification}
\label{sub:model-uq}

We provide built-in machine-learning uncertainties for our models using the last-layer prediction rigidity (LLPR) method \cite{bigi2024prediction}. Within the LLPR formalism, the uncertainty of the model's prediction in total energy is computed as
\begin{equation}
    \sigma^2 = \alpha^2 \boldsymbol{x}^\top (\boldsymbol{X}^\top \boldsymbol{X} + \varsigma^2 \boldsymbol{I})^{-1} \boldsymbol{x},
\end{equation}
where $\sigma$ is the predictive uncertainty, $\boldsymbol{x}$ are the last-layer features associated to the predicted sample, $\boldsymbol{X}$ is a matrix whose rows correspond to the last-layer features of all samples in the training set, $\varsigma^2$ is a small positive regularization parameter, and the calibration factor $\alpha^2$ is determined on a calibration set. 
Given that the last-layer features have to be computed and materialized during the model's forward pass, this uncertainty quantification method adds negligible overhead to its predictions. Beyond predictive uncertainties, this construction allows users to compute a range of local and global rigidity measures~\cite{chong2023robustness,chong2025prediction}. 
The LLPR covariance matrix $\boldsymbol{X}^\top \boldsymbol{X}$ is also used to sample the members of a last-layer ensemble as described in Ref.~\cite{bigi2024prediction}. This further allows the propagation of uncertainties through arbitrarily complex workflows~\cite{kellner2024uncertainty,mazitov2025pet} when using our models.

  \begin{table*}
      \centering
      \begin{tabular*}{\textwidth}{@{\extracolsep{\fill}}lcccccc}
      \toprule
       & \multicolumn{2}{c}{PET-MAD-1.6-XS} & \multicolumn{2}{c}{PET-MAD-1.6-S} & \multicolumn{2}{c}{PET-MAD-1.6-M} \\
      \cmidrule(lr){2-3} \cmidrule(lr){4-5} \cmidrule(lr){6-7}
     Subset & $E$ & ${\mathbf{f}}$ & $E$ & ${\mathbf{f}}$ & $E$ & ${\mathbf{f}}$ \\
      \midrule
      MC3D & 10.72 & 53.64 & 4.82 & 26.67 & 3.09 & 17.67 \\
      MC3D-rattled & 10.82 & 137.36 & 5.21 & 76.78 & 3.58 & 49.37 \\
      MC3D-random & 49.49 & 154.74 & 34.38 & 99.94 & 33.48 & 76.42 \\
      MC3D-surface & 11.82 & 96.04 & 5.48 & 54.45 & 4.04 & 35.78 \\
      MC3D-cluster & 21.81 & 126.45 & 12.68 & 76.43 & 10.41 & 57.57 \\
      MC2D & 13.96 & 66.15 & 6.14 & 34.89 & 4.79 & 22.78 \\
      SHIFTML-molcrys & 3.76 & 53.66 & 1.47 & 24.49 & 0.83 & 15.24 \\
      SHIFTML-molfrags & 5.41 & 54.83 & 2.12 & 25.06 & 1.18 & 17.10 \\
      \midrule
      MC3D-extended & 17.49 & 58.73 & 12.38 & 33.43 & 10.04 & 23.87 \\
      binary-random & 24.83 & 31.12 & 18.06 & 19.50 & 15.86 & 14.99 \\
      MC3D-random-ext & 44.53 & 187.49 & 31.29 & 126.64 & 31.54 & 104.22 \\
      MAD-cat & 3.27 & 69.58 & 1.48 & 43.15 & 0.99 & 32.24 \\
      \midrule
      Total & 15.01 & 71.64 & 9.62 & 41.65 & 8.10 & 29.73 \\
      \bottomrule
      \end{tabular*}
      \caption{Performance of the PET-MAD-1.6 models on the subset-resolved MAD-1.6 test set. Mean absolute error (MAE) in energy
  $|$ forces prediction is given in meV/atom $|$ meV/\AA. }
      \label{tab:mad-test-results}
  \end{table*}

\begin{figure*}[t]
    \centering
    \includegraphics[width=1.0\textwidth]{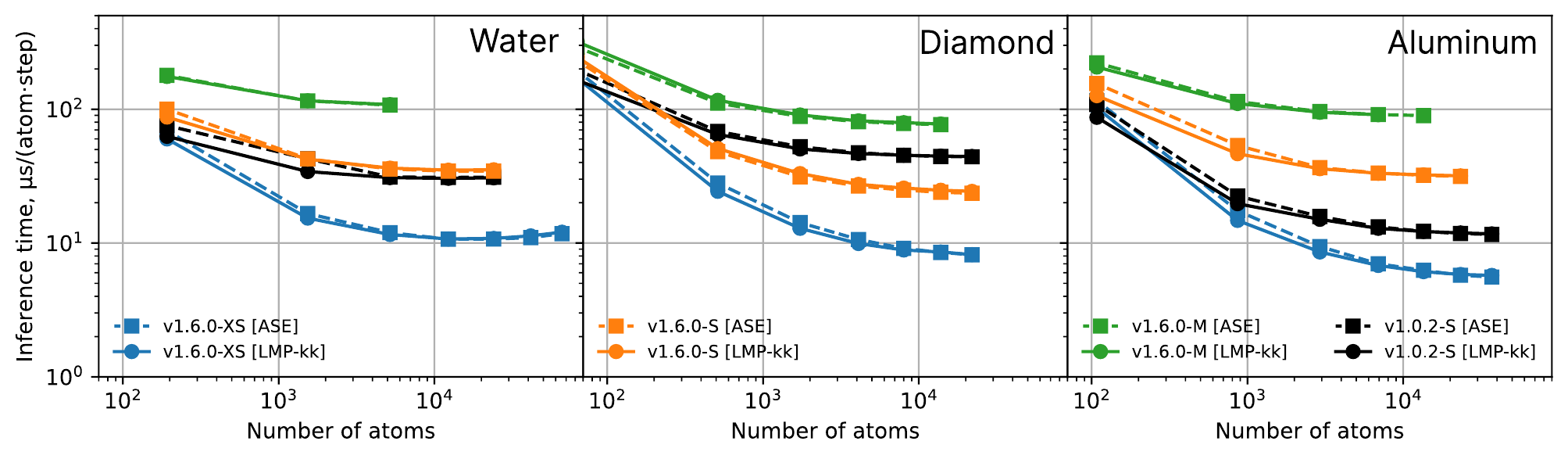}
    \label{fig:speed}
    \caption{Inference time of PET-MAD universal models evaluated over different kinds of materials and varying system sizes. The results of the XS-, S- and M-size PET-MAD-1.6 models is shown with blue and orange lines, respectively, where circle and square markers are denoting the ASE \cite{ase} and LAMMPS (with Kokkos backend) \cite{lammps, kokkos} implementations. The original PET-MAD-1 timings with both the model weights and code implementations from Ref.~\citenum{mazitov2025pet} are shown with gray lines. The PET-MAD-1 timings with the updated code implementation associated with the PET-MAD-1.6 release are shown in black. All performance test were done on a single NVIDIA H100 GPU. Conservative heads were used for all the predictions. 
    }
\end{figure*}

\subsection{Benchmarking}

Since MAD-1.6 consists of a diverse set of material classes, with systems that span different sizes, dimensionalities and chemistries, its performance on a test subset can serve a first, simple measure of its expected accuracy for different types of problems. 
A subset-resolved analysis of each model's errors is shown in Table \ref{tab:mad-test-results}. As expected, the S-size model achieves much greater accuracy than the XS model, with an excellent mean absolute force error (MAE) value of 42 meV/\AA on the MAD-1.6 test set. 
Even though this value is dominated by large and low-error subsets like \emph{MC3D} or \emph{binary-random}, is still does not exceed 77 meV/\AA \ for all ``reasonable'' configurations, and only reaches 127 meV/\AA \ on the \emph{MC3D-random-extended} subset. 
The M-size model reduces errors even further, surpassing the 30 meV/\AA \ threshold on the overall test set and achieving an unparalleled range of 15 to 58 meV/\AA \ on the ``reasonable'' configurations, while ultimately not exceeding 104 meV/\AA \ on the remaining subsets.
Although the XS model has a roughly five times smaller number of parameters, it still achieves a level of accuracy that is comparable to that of the original PET-MAD-1, while covering a larger chemical space and having up to 3 times lower inference time compared to its S-size analog. Figure~\ref{fig:speed} shows timings of the models for a few different kinds of materials in both the ASE \cite{ase} and LAMMPS \cite{lammps} implementations, while also including a comparison with the original implementation used in PET-MAD-1. 

\begin{table}
    \centering
    \begin{tabular*}{\columnwidth}{@{\extracolsep{\fill}}lccccccccc}
    \toprule
     & \multicolumn{2}{c}{v1.6-XS} & \multicolumn{2}{c}{v1.6-S} & \multicolumn{2}{c}{v1.6-M} & \multicolumn{2}{c}{v1.0.2} \\
     & \multicolumn{2}{c}{(this work)} & \multicolumn{2}{c}{(this work)}& \multicolumn{2}{c}{(this work)}& \multicolumn{2}{c}{(Ref.~\citenum{mazitov2025pet})} \\
   Subset & $E$ & ${\mathbf{f}}$  & $E$ & ${\mathbf{f}}$  &  $E$ & ${\mathbf{f}}$  & $E$ & ${\mathbf{f}}$ \\
    \midrule
    MAD-1 & 9.59 & 71.35 & 4.31 & 34.58 & \textbf{2.76} & \textbf{20.26} & 17.6  &  65.1 \\
    MPtrj & 15.61 & 84.84 & 9.74 & 55.72 & \textbf{7.61} & \textbf{40.98} & 22.3  &  77.9\\
    MatBench & 30.00 & 63.40 & 17.56 & 43.92 & \textbf{12.78} & \textbf{33.45} & 31.3  &  — \\
    Alexandria & 39.53 & 60.30 & 25.79 & 44.81 & \textbf{20.51} & \textbf{33.74} & 49.0  &  66.8\\
    OC2020 & 14.15 & 111.41 & 8.08 & 76.63 & \textbf{6.85} & \textbf{60.89} &  18.3  &  114.5\\
    SPICE & 4.06 & 68.21 & 4.15 & 61.42 & \textbf{2.13} & \textbf{31.61} & 3.7  &  59.5\\
    MD22 & 3.77 & 81.70 & 5.08 & 68.80 & 2.05 & \textbf{36.76 }& \textbf{1.9}  &  65.6\\
    \bottomrule
    \end{tabular*}
    \caption{Performance of the PET-MAD-1.6 models on the subset-resolved MADBench benchmark. Mean absolute error (MAE) in energy and forces prediction is given in meV/atom and meV/\AA, respectively. Results of the PET-MAD v1.0.2 trained on the original PBEsol version of the MAD-1 dataset from Ref.~\citenum{mazitov2025pet} are provided for comparison. Each model is evaluated against a set of consistently computed DFT targets corresponding to those used for training.  The results of the best model for each subset are highlighted in bold. }
    \label{tab:madbench-results}
\end{table}

In addition to the in-domain evaluation on a test subset, we performed an out-of-domain (OOD) analysis of the models' performance using a MADBench benchmark - a minimalistic set of configurations sampled from various datasets used for atomistic machine learning from Ref.~\citenum{mazitov2025pet}. The main advantage of this benchmark is the availability of targets computed consistently with various DFT settings. This allows for direct comparison of the models, ensuring that discrepancies are not caused by significant differences in the reference data. We recomputed the target energies and forces on the MADBench configurations using the same DFT setup as for the MAD-1.6 calculations, and evaluated the PET-MAD-1.6 models against them. The subset-resolved analysis of the models' errors is provided in Table \ref{tab:madbench-results}. 
Even though only the MAD subset of the MADBench can be considered as a pure in-domain part of the benchmark, both XS- and S-size models show good accuracies on the other OOD subsets, while the latter gets a substantial improvement over the original PET-MAD-1 model. Similar to the evaluation on a test subset, the errors on forces do not exceed 77 meV/\AA \ and 61 meV/\AA \ for the S and M models, respectively, indicating the models' excellent generalizability. 

We note, that these results are unprecedented, especially given that our models (1) cover a very large chemical space and (2) target the r$^2$SCAN level of theory. 
Massive datasets of several million structures are practically unavailable for this level of theory due to the increased cost and lower robustness of meta-GGA DFT.
The high accuracy of the models we propose is striking when compared, for instance, to the largest models trained on OMat24 (which achieve around 45-50 meV/\AA ~\cite{bigi2026pushing} on a massive PBE dataset), MAD-1 (around 70-75~meV/\AA ~\cite{mazitov2025pet}, PBEsol), or MATPES ($>$100~meV/\AA ~\cite{matpes}, both in its PBE and r$^2$SCAN versions).
Compared to these test subset results, our models achieve a MAE of 72 meV/\AA, 42 meV/\AA, and 30 meV/\AA  \ in forces prediction for the XS, S, and M sizes, respectively. Covering such a large chemical space at a high level of theory and accuracy is made possible by predominantly two facts: first, our models are pre-trained on the extensive OMat24 dataset, which allows for the creation of high-quality neural representations. 
Second, they are fine-tuned on the smaller, but accurate and carefully curated MAD-1.6 dataset, which maintains a high level of internal consistency in the underlying DFT, and therefore avoids unlearnable noise in target data that would deteriorate the models' accuracy.
Last, but not the least, the new models are fast, with S-sized model being comparable to PET-MAD-1.0, XS model being much faster, and M model being only about 2 to 2.5 times slower compared to the S-size in favor of higher accuracy (Fig.~\ref{fig:speed}).


\subsection{Mendeleev clusters}

As a challenging test of the stability of the potential, we set up the following benchmark simulations, inspired by similar simulations performed by some of us for high-entropy alloys~\cite{lopa+23prm,mazi+24jpm}.
We generate ``Mendeleev clusters'' by creating a $3\times 3\times 3$ \emph{fcc} box, and assign at random to each site one of the atoms in the periodic table, leaving six vacancies to avoid repetitions. The initial structure is relaxed, keeping the cell cubic, and the boundary conditions then extended to a side of 40~\AA{}, effectively creating an initial particle with a cubic shape. 
This procedure is repeated to create 16 structures, each of which is used as the initial configuration in a replica exchange molecular dynamics (REMD) simulation, with replicas at temperatures between 300~K and 3000~K, uniformly spaced on a logarithmic scale, using the implementation in i-PI~\cite{petr+15jcc,litm+24jcp} and the PET-MAD-1.5-S model.
We run 100~ps-long trajectories in the \emph{NVT} ensemble, using an efficient thermostat combining a stochastic velocity rescaling\cite{buss+07jcp} and an optimal-sampling Generalized Langevin Equation thermostat\cite{ceri+10jctc}. 
We also attempt, every two steps, a random exchange between the position of two atoms, to accelerate the exploration of configuration space.
\begin{figure}[tb]
    \centering    \includegraphics[width=1.0\linewidth]{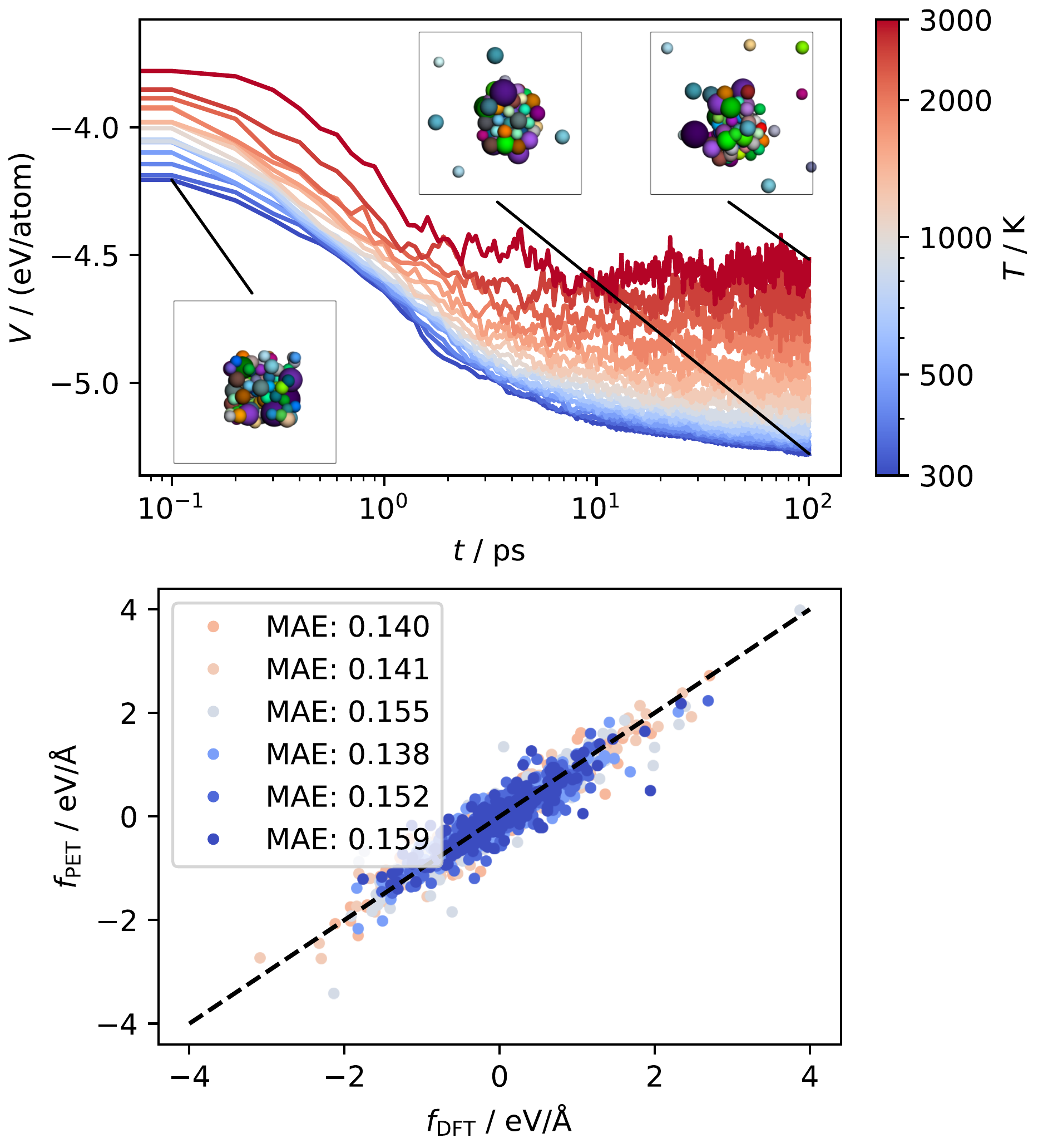}
    \caption{(top) Time series of the potential energy for the re-ordered temperature replicas for a REMD simulation of a Mendeleev cluster containing 102 elements. 
    Insets show the initial and final structures of the 300~K replica, and the final structure of the 3000~K replica. Isolated atoms in the end of the simulation are typically either noble gases or elements with a high vapor pressure (such as Hg).
    (bottom) Parity plot of the force components for the final structures in the simulation, color-coded based on the target temperature, comparing the values obtained with PET-MAD-1.6-M with those computed with single-point DFT calculations. Typical errors are of the order of 150 meV/\AA{} MAE. 
    }
    \label{fig:mendeleev}
\end{figure}

This is an incredibly challenging setup, probing high temperatures, atoms in bulk, surface and (potentially) gas phase environments, and efficient sampling to maximize the possibility of encountering unphysical low-energy minima. 
The potential generates stable trajectories, resulting in a compact quasi-spherical particle that has expelled all and only the noble gases for the low-temperature replicas, while at the highest temperature the surface of the cluster becomes irregular, and one can find occasionally isolated atoms of other elements, as well as alkali halide dimers  (Figure~\ref{fig:mendeleev}). The refined PET-MAD-1.6 potential reproduces this behavior, yielding equally stable simulations and equivalent structural outcomes.
To provide a more quantitative assessment of the reliability of the potential in this extreme regime, we performed single-point r$^2$SCAN calculations for some of the final structures obtained from the PET-MAD-1.5 REMD trajectory. 
The convergence of these calculations is extremely challenging, requiring careful tuning of the self-consistency loop and a relaxation of the convergence threshold. 
Even though this indicates that the DFT reference may itself be affected by a degree of uncertainty, the comparison between the electronic-structure calculations the predictions are reassuring, with a force error around 150 meV/\AA{} when using PET-MAD-1.6-M (Figure~\ref{fig:mendeleev}, bottom - similar errors are obtained also with PET-MAD-1.5, and with the lighter S models), in line with the validation error on the \emph{MC3D-random-extended} subset, and lower than the errors that first-generation MLIPs would yield for single-component bulk materials~\cite{behl-parr07prl}.

\section{Conclusions}

We present MAD-1.6, a dataset designed to cover the entire periodic table up to \ce{No}, which includes all isotopes with a half-life above a day, covering organic and inorganic materials, bulk structures, surfaces, clusters and molecules.
In line with the design principle of ``massive atomic diversity'', MAD-1.6 contains about 360{,}000 structures, chosen to be representative of realistic materials and molecules, but to also include very high-energy artificial structures.

We use reference calculations that are highly converged, based on the all-electron numeric atomic orbitals code FHI-aims, and use the r$^2$SCAN meta-GGA functional to strike a good balance between computational cost and accuracy. 
We further improve the internal consistency of MAD-1.6 by performing an outlier screening procedure based on the calculation of last-layer rigidity uncertainties. We also make available the structures that are discarded for having a very large empirical error-to-uncertainty ratio, as they are likely to represent hard-to-converge benchmarks for DFT calculations.

Based on the cleaned MAD-1.6 dataset, we train demonstrative models based on the unconstrained PET architecture. PET-MAD-1.6-XS, PET-MAD-1.6-S and PET-MAD-1.6-M achieve excellent performance compared to models that are practically usable for advanced materials simulations, while covering a much larger portion of chemical space. 
Besides being accurate in benchmarks, PET-MAD-1.6 models are also extremely stable. We introduce a ``Mendeleev cluster'' stress test that consists in simulating a nanoparticle containing one each of all the elements present in MAD-1.6 across a broad range of temperatures. The simulation remains stable over a collective 1.6~ns of trajectory, and lead to physically plausible outcomes and an error of about 150 meV/\AA{} relative to explicit electronic-structure calculations. 
This is possibly the most compelling demonstration of the universal applicability of PET-MAD-1.6, that in turns showcases the effectiveness of a dataset construction strategy focused on diversity and internal consistency of the electronic-structure calculations.

\section{Data record and model availability}

The MAD-1.6 dataset is made publicly available as a record~\cite{MAD-1.6-mca} within the Materials Cloud Archive~\cite{MaterialsCloud}, which is a FAIR repository dedicated to materials-science simulations. Structures and targets computed at r$^2$SCAN and PBE levels of theory that comprise the dataset are provided in extended XYZ format. Energies (specifically total energy under key \texttt{energy}, and atomization energy under key \texttt{atomization\_energy}), stresses (\texttt{stress}), and lattice parameters (\texttt{Lattice}) are stored in the file headers, and atom types (\texttt{species}), positions (\texttt{pos}), and forces (\texttt{forces}) are stored as space-separated entries. Cartesian coordinates, energies, forces, and stresses are given in \si{\angstrom}, \si{\electronvolt}, \si{\electronvolt}/\si{\angstrom}, and \si{\electronvolt}/\si{\angstrom}$^3$, respectively. Also stored in the header are the name of the subset to which each structure belongs (\texttt{subset}) and the numeric index of the structure within its subset (\texttt{frame\_id}). Together these uniquely identify each structure in the dataset. We note that for the r$^2$SCAN data, stresses are included for periodic structures from the \textit{MC3D-rattled} (\texttt{mc3d\_rattled}) subset only.

Specifically, the record contains:
\begin{itemize}
    \item \textit{mad-1.6-r2scan-train.xyz} - the training set used in model training.
    \item \textit{mad-1.6-r2scan-val.xyz} - the validation set used in model training.
    \item \textit{mad-1.6-r2scan-test.xyz} - the test set used in model evaluation.
    \item \textit{mad-1.6-r2scan-llpr-rejected.xyz} - the structures removed by the LLPR uncertainty-based filtering procedure shown in Fig.~\ref{fig:llpr-cleaning}.
    \item \textit{mad-1.6-pbe.xyz} - the combined training, validation, and test splits (consistent with the r$^2$SCAN splits) of a subset of MAD-1.6 computed with the PBE functional, used in model training.
\end{itemize}

We also release PET-MAD-1.6 as three foundation MLIPs with varying number of parameters, trained on the above data splits. These are available in the public GitHub repository \url{https://github.com/lab-cosmo/upet}.
An example to reproduce the simulation of the Mendeleev cluster with these (and other) universal MLIPs can be found at \url{https://atomistic-cookbook.org/examples/mendeleev/mendeleev.html}.

\section{Author contributions}
C.M. performed all the electronic-structure calculations for the MAD-1.6 dataset, generated the \emph{MC3D-random-extended} dataset, the relaxed part of the \emph{MC3D-extended} subset and the  \emph{trimer-extended} subset. C.M. and P.L. generated the \emph{monomers}, \emph{dimers}, and \emph{trimers} subsets. P.P. generated the \emph{binary-random} subset. T.J.G. and S.D. developed and executed the strategy for the training set configurations of \emph{MAD-cat} subset.
A.M. trained the PET-MAD-1.6-XS, PET-MAD-1.6-S and PET-MAD-1.6-M models, carried out the uncertainty quantification protocol used to identify outliers, ran the accuracy and speed benchmark evaluations, and managed the model distribution. F.B. trained preliminary models and provided support in model training. A.M. and F.B. managed the UPET repository and PET model code implementation. 
J.W.A. provided technical support for the initial setup of the electronic-structure calculations and curated the data distribution on Materials Cloud. M.R. and C.M. addressed convergence issues for structures containing d- and f-block elements. M.R. provided technical support for the electronic-structure calculations and performed the electronic-structure calculations for the Mendeleev cluster.
A.M. organized and managed the technical part of the project. M.C. supervised and guided the project and performed the REMD simulations.
All authors contributed to the writing and revision of the manuscript.

\begin{acknowledgments}
The Authors are grateful to Damien K. J. Lee and Anirudh Raju Natarajan for providing template orderings of the crystal structures used for the \emph{binary-random} subset generation and to Marnik Bercx, Sebastian Huber and Giovanni Pizzi for providing the initial configurations for the \emph{MC3D-extended} subset of MAD-1.5.
A.M. was supported by an Industrial Grant from BASF. P.L., P.P. and M.C. acknowledge support by the NCCR MARVEL, funded by the Swiss National Science Foundation (SNSF, grant number 182892). F.B. was supported by a grant from the Platform for Advanced Scientific Computing (PASC) and the Swiss AI Initiative (2025 Fellowship Program). J.W.A. and M.C. received funding from the European Research Council (ERC) under the European Union’s Horizon 2020 research and innovation programme (grant agreement No. 101001890 - FIAMMA).
M.R. and C.M. acknowledge computer time at the Max Planck Computing and Data Facility (MPCDF). C.M., F.B., P.P., J.W.A., P.L., M.C. and A.M. acknowledge computational resources at the Swiss National Supercomputing Center, Switzerland with project ID lp133 on Daint and ID mr31 under MARVEL's share on Eiger, the SwissAI initiative projects aa002 and a117 as well as the SCITAS platform at EPFL.
\end{acknowledgments}

\bibliography{others,biblio}

@article{AIMS,
  title = {Ab Initio Molecular Simulations with Numeric Atom-Centered Orbitals},
  author = {Blum, Volker and Gehrke, Ralf and Hanke, Felix and Havu, Paula and Havu, Ville and Ren, Xinguo and Reuter, Karsten and Scheffler, Matthias},
  year = 2009,
  month = nov,
  journal = {Computer Physics Communications},
  volume = {180},
  number = {11},
  pages = {2175--2196},
  publisher = {Elsevier B.V.},
  issn = {00104655},
  doi = {10.1016/j.cpc.2009.06.022},
  urldate = {2013-01-31}
}

@article{bart+13prb,
  title = {On Representing Chemical Environments},
  author = {Bart{\'o}k, Albert P. and Kondor, Risi and Cs{\'a}nyi, G{\'a}bor},
  year = 2013,
  month = may,
  journal = {Physical Review B},
  volume = {87},
  number = {18},
  pages = {184115},
  issn = {1098-0121},
  doi = {10.1103/PhysRevB.87.184115},
  urldate = {2014-09-24}
}

@article{behl-parr07prl,
  title = {Generalized {{Neural-Network Representation}} of {{High-Dimensional Potential-Energy Surfaces}}},
  author = {Behler, J{\"o}rg and Parrinello, Michele},
  year = 2007,
  month = apr,
  journal = {Physical Review Letters},
  volume = {98},
  number = {14},
  pages = {146401},
  issn = {0031-9007},
  doi = {10.1103/PhysRevLett.98.146401},
  urldate = {2013-11-11}
}

@inproceedings{bigi+25icml,
  title = {The Dark Side of the Forces: Assessing Non-Conservative Force Models for Atomistic Machine Learning},
  booktitle = {Forty-Second International Conference on Machine Learning},
  author = {Bigi, Filippo and Langer, Marcel F. and Ceriotti, Michele},
  year = 2025
}

@article{buss+07jcp,
  title = {Canonical Sampling through Velocity Rescaling},
  author = {Bussi, G and Donadio, D and Parrinello, M},
  year = 2007,
  journal = {Journal of Chemical Physics},
  volume = {126},
  number = {1},
  pages = {14101}
}

@article{ceri+10jctc,
  title = {Colored-Noise Thermostats \`a La {{Carte}}},
  author = {Ceriotti, Michele and Bussi, Giovanni and Parrinello, Michele},
  year = 2010,
  month = apr,
  journal = {Journal of Chemical Theory and Computation},
  volume = {6},
  number = {4},
  pages = {1170--1180},
  publisher = {ACS Publications},
  issn = {15499626},
  doi = {10.1021/ct900563s},
  urldate = {2013-02-08}
}

@article{chong+26jcp,
  title = {Resolving the Body-Order Paradox of Machine Learning Interatomic Potentials},
  author = {Chong, Sanggyu and Jiang, Tong and Domina, Michelangelo and Bigi, Filippo and Grasselli, Federico and Lee, Joonho and Ceriotti, Michele},
  year = 2026,
  month = feb,
  journal = {The Journal of Chemical Physics},
  volume = {164},
  number = {6},
  pages = {064121},
  issn = {0021-9606, 1089-7690},
  doi = {10.1063/5.0303302},
  urldate = {2026-02-14},
  langid = {english}
}

@article{cohe+08science,
  title = {Insights into Current Limitations of Density Functional Theory.},
  author = {Cohen, Aron J and {Mori-S{\'a}nchez}, Paula and Yang, Weitao},
  year = 2008,
  month = aug,
  journal = {Science},
  volume = {321},
  number = {5890},
  eprint = {18687952},
  eprinttype = {pubmed},
  pages = {792--4},
  issn = {1095-9203},
  doi = {10.1126/science.1158722},
  urldate = {2013-05-21},
  pmid = {18687952}
}

@article{de+16pccp,
  title = {Comparing Molecules and Solids across Structural and Alchemical Space},
  author = {De, Sandip and Bart{\'o}k, Albert P. and Cs{\'a}nyi, G{\'a}bor and Ceriotti, Michele},
  year = 2016,
  journal = {Physical Chemistry Chemical Physics},
  volume = {18},
  number = {20},
  eprint = {1601.04077},
  pages = {13754--13769},
  publisher = {The Royal Society of Chemistry},
  issn = {14639076},
  doi = {10.1039/c6cp00415f},
  archiveprefix = {arXiv}
}

@article{jage+18npjcm,
  title = {Machine Learning Hydrogen Adsorption on Nanoclusters through Structural Descriptors},
  author = {J{\"a}ger, Marc O. J. and Morooka, Eiaki V. and Federici Canova, Filippo and Himanen, Lauri and Foster, Adam S.},
  year = 2018,
  month = dec,
  journal = {npj Comput Mater},
  volume = {4},
  number = {1},
  pages = {37},
  issn = {2057-3960},
  doi = {10.1038/s41524-018-0096-5},
  urldate = {2021-04-15},
  langid = {english}
}

@article{litm+24jcp,
  title = {I-{{PI}} 3.0: {{A}} Flexible and Efficient Framework for Advanced Atomistic Simulations},
  shorttitle = {I-{{PI}} 3.0},
  author = {Litman, Yair and Kapil, Venkat and Feldman, Yotam M. Y. and Tisi, Davide and Begu{\v s}i{\'c}, Tomislav and Fidanyan, Karen and Fraux, Guillaume and Higer, Jacob and Kellner, Matthias and Li, Tao E. and P{\'o}s, Eszter S. and Stocco, Elia and Trenins, George and Hirshberg, Barak and Rossi, Mariana and Ceriotti, Michele},
  year = 2024,
  month = aug,
  journal = {The Journal of Chemical Physics},
  volume = {161},
  number = {6},
  pages = {062504},
  issn = {0021-9606, 1089-7690},
  doi = {10.1063/5.0215869},
  urldate = {2024-08-16},
  langid = {english}
}

@article{lopa+23prm,
  title = {Modeling High-Entropy Transition Metal Alloys with Alchemical Compression},
  author = {Lopanitsyna, Nataliya and Fraux, Guillaume and Springer, Maximilian A. and De, Sandip and Ceriotti, Michele},
  year = 2023,
  month = apr,
  journal = {Phys. Rev. Materials},
  volume = {7},
  number = {4},
  pages = {045802},
  issn = {2475-9953},
  doi = {10.1103/PhysRevMaterials.7.045802},
  urldate = {2023-04-27},
  langid = {english}
}

@article{mazi+24jpm,
  title = {Surface Segregation in High-Entropy Alloys from Alchemical Machine Learning},
  author = {Mazitov, Arslan and Springer, Maximilian A and Lopanitsyna, Nataliya and Fraux, Guillaume and De, Sandip and Ceriotti, Michele},
  year = 2024,
  month = apr,
  journal = {J. Phys. Mater.},
  volume = {7},
  number = {2},
  pages = {025007},
  issn = {2515-7639},
  doi = {10.1088/2515-7639/ad2983},
  urldate = {2024-03-01}
}

@article{mazi+25ncomm,
  title = {{{PET-MAD}} as a Lightweight Universal Interatomic Potential for Advanced Materials Modeling},
  author = {Mazitov, Arslan and Bigi, Filippo and Kellner, Matthias and Pegolo, Paolo and Tisi, Davide and Fraux, Guillaume and Pozdnyakov, Sergey and Loche, Philip and Ceriotti, Michele},
  year = 2025,
  month = nov,
  journal = {Nat Commun},
  volume = {16},
  number = {1},
  pages = {10653},
  issn = {2041-1723},
  doi = {10.1038/s41467-025-65662-7},
  urldate = {2025-12-09},
  langid = {english}
}

@article{mazi+25sd,
  title = {Massive {{Atomic Diversity}}: A Compact Universal Dataset for Atomistic Machine Learning},
  shorttitle = {Massive {{Atomic Diversity}}},
  author = {Mazitov, Arslan and Chorna, Sofiia and Fraux, Guillaume and Bercx, Marnik and Pizzi, Giovanni and De, Sandip and Ceriotti, Michele},
  year = 2025,
  month = nov,
  journal = {Sci Data},
  volume = {12},
  number = {1},
  pages = {1857},
  issn = {2052-4463},
  doi = {10.1038/s41597-025-06109-y},
  urldate = {2025-11-21},
  langid = {english}
}

@article{paru+18ncomm,
  title = {Chemical Shifts in Molecular Solids by Machine Learning},
  author = {Paruzzo, Federico M. and Hofstetter, Albert and Musil, F{\'e}lix and De, Sandip and Ceriotti, Michele and Emsley, Lyndon},
  year = 2018,
  month = dec,
  journal = {Nature Communications},
  volume = {9},
  number = {1},
  eprint = {1805.11541},
  pages = {4501},
  publisher = {Springer US},
  issn = {20411723},
  doi = {10.1038/s41467-018-06972-x},
  archiveprefix = {arXiv}
}

@incollection{pasz+19nips,
  title = {{{PyTorch}}: {{An}} Imperative Style, High-Performance Deep Learning Library},
  booktitle = {Advances in Neural Information Processing Systems 32},
  author = {Paszke, Adam and Gross, Sam and Massa, Francisco and Lerer, Adam and Bradbury, James and Chanan, Gregory and Killeen, Trevor and Lin, Zeming and Gimelshein, Natalia and Antiga, Luca and Desmaison, Alban and Kopf, Andreas and Yang, Edward and DeVito, Zachary and Raison, Martin and Tejani, Alykhan and Chilamkurthy, Sasank and Steiner, Benoit and Fang, Lu and Bai, Junjie and Chintala, Soumith},
  editor = {Wallach, H. and Larochelle, H. and Beygelzimer, A. and {dAlch{\'e}-Buc}, F. and Fox, E. and Garnett, R.},
  year = 2019,
  pages = {8024--8035},
  publisher = {Curran Associates, Inc.}
}

@article{perd+08prl,
  title = {Restoring the {{Density-Gradient Expansion}} for {{Exchange}} in {{Solids}} and {{Surfaces}}},
  author = {Perdew, John P. and Ruzsinszky, Adrienn and Csonka, G{\'a}bor I. and Vydrov, Oleg A. and Scuseria, Gustavo E. and Constantin, Lucian A. and Zhou, Xiaolan and Burke, Kieron},
  year = 2008,
  month = apr,
  journal = {Phys. Rev. Lett.},
  volume = {100},
  number = {13},
  pages = {136406},
  issn = {0031-9007, 1079-7114},
  doi = {10.1103/PhysRevLett.100.136406},
  urldate = {2022-11-25},
  langid = {english}
}

@article{petr+15jcc,
  title = {Beyond Static Structures: {{Putting}} Forth {{REMD}} as a Tool to Solve Problems in Computational Organic Chemistry},
  author = {Petraglia, Riccardo and Nicola{\"i}, Adrien and Wodrich, Matthew D. and Ceriotti, Michele and Corminboeuf, Clemence},
  year = 2016,
  month = jan,
  journal = {Journal of Computational Chemistry},
  volume = {37},
  number = {1},
  pages = {83--92},
  issn = {1096987X},
  doi = {10.1002/jcc.24025}
}

@article{will+18pccp,
  title = {Feature Optimization for Atomistic Machine Learning Yields a Data-Driven Construction of the Periodic Table of the Elements},
  author = {Willatt, Michael J. and Musil, F{\'e}lix and Ceriotti, Michele},
  year = 2018,
  journal = {Physical Chemistry Chemical Physics},
  volume = {20},
  number = {47},
  pages = {29661--29668},
  publisher = {Royal Society of Chemistry},
  issn = {14639076},
  doi = {10.1039/c8cp05921g}
}

@article{arpess,
  author  = {Plessow, Philipp N.},
  title   = {Efficient Transition State Optimization of Periodic Structures through Automated Relaxed Potential Energy Surface Scans},
  journal = {J. Chem. Theory Comput.},
  year    = 2018,
  month   = January,
  volume = {14},
  number = {2},
  pages = {981-990},
  doi     = {10.1021/acs.jctc.7b01070}
}

@article{autoadsorbate,
  author  = {Fako, Edvin and De, Sandip},
  title   = {Simple Heuristics for Advanced Sampling of Reactive Species on Surfaces},
  journal = {ACS Catal.},
  volume = {16},
  number = {4},
  pages = {3149-3158},
  year = {2026},
  month   = February,
  volume = {14},
  number = {2},
  pages = {981-990},
  doi     = {10.1021/acscatal.5c06553}
}

@misc{MAD-1.6-mca,
  doi = {10.24435/materialscloud:jc-9f},
  url = {https://archive.materialscloud.org/doi/10.24435/materialscloud:jc-9f},
  author = {Malosso, Cesare and Bigi, Filippo and Pegolo, Paolo and Abbott, Joseph W. and Loche, Philip and Rossi, Mariana and Ceriotti, Michele and Mazitov, Arslan},
  title = {A high-quality, high-information dataset for universal atomistic machine learning},
  howpublished={DOI: 10.24435/materialscloud:jc-9f},
  publisher = {Materials Cloud},
  year = {2026},
  copyright = {Creative Commons Attribution 4.0 International}
}

@article{MC3D,
  title = {MC3D: the materials cloud computational database of experimentally known stoichiometric inorganics},
  ISSN = {2635-098X},
  url = {http://dx.doi.org/10.1039/D5DD00415B},
  DOI = {10.1039/d5dd00415b},
  journal = {Digital Discovery},
  publisher = {Royal Society of Chemistry (RSC)},
  author = {Huber,  Sebastiaan P. and Minotakis,  Michail and Bercx,  Marnik and Reents,  Timo and Eimre,  Kristjan and Paulish,  Nataliya and H\"{o}rmann,  Nicolas and Uhrin,  Martin and Marzari,  Nicola and Pizzi,  Giovanni},
  year = {2026}
}

@misc{FHI-aims2025,
  doi = {10.48550/ARXIV.2505.00125},
  url = {https://arxiv.org/abs/2505.00125},
  author = {Abbott, Joseph W. and Acosta, Carlos Mera and Akkoush, Alaa and Ambrosetti, Alberto and Atalla, Viktor and Bagrets, Alexej and Behler, J\"{o}rg and Berger, Daniel and Bieniek, Bj\"{o}rn and Bj\"{o}rk, Jonas and Blum, Volker and Bohloul, Saeed and Box, Connor L. and Boyer, Nicholas and Brambila, Danilo Simoes and Bramley, Gabriel A. and Bryenton, Kyle R. and Camarasa-G\'{o}mez, Mar\'{i}a and Carbogno, Christian and Caruso, Fabio and Chutia, Sucismita and Ceriotti, Michele and Cs\'{a}nyi, G\'{a}bor and Dawson, William and Delesma, Francisco A. and Della Sala, Fabio and Delley, Bernard and DiStasio, Jr., Robert A. and Dragoumi, Maria and Driessen, Sander and Dvorak, Marc and Erker, Simon and Evers, Ferdinand and Fabiano, Eduardo and Farrow, Matthew R. and Fiebig, Florian and Filser, Jakob and Foppa, Lucas and Gallandi, Lukas and Garcia, Alberto and Gehrke, Ralf and Ghan, Simiam and Ghiringhelli, Luca M. and Glass, Mark and Goedecker, Stefan and Golze, Dorothea and Gramzow, Matthias and Green, James A. and Grisafi, Andrea and Gr\"{u}neis, Andreas and G\"{u}nzl, Jan and Gutzeit, Stefan and Hall, Samuel J. and Hanke, Felix and Havu, Ville and He, Xingtao and Hekele, Joscha and Hellman, Olle and Herath, Uthpala and Hermann, Jan and Hernang\'{o}mez-P\'{e}rez, Daniel and Hofmann, Oliver T. and Hoja, Johannes and Hollweger, Simon and H\"{o}rmann, Lukas and Hourahine, Ben and How, Wei Bin and Huhn, William P. and H\"{u}lsberg, Marcel and Jacob, Timo and Jand, Sara Panahian and Jiang, Hong and Johnson, Erin R. and J\"{u}rgens, Werner and Kahk, J. Matthias and Kanai, Yosuke and Kang, Kisung and Karpov, Petr and Keller, Elisabeth and Kempt, Roman and Khan, Danish and Kick, Matthias and Klein, Benedikt P. and Kloppenburg, Jan and Knoll, Alexander and Knoop, Florian and Knuth, Franz and K\"{o}cher, Simone S. and Kockl\"{a}uner, Jannis and Kokott, Sebastian and K\"{o}rzd\"{o}rfer, Thomas and Kowalski, Hagen-Henrik and Kratzer, Peter and K\r{u}s, Pavel and Laasner, Raul and Lang, Bruno and Lange, Bj\"{o}rn and Langer, Marcel F. and Larsen, Ask Hjorth and Lederer, Hermann and Lehtola, Susi and Lenz-Himmer, Maja-Olivia and Leucke, Moritz and Levchenko, Sergey and Lewis, Alan and {von Lilienfeld}, O. Anatole and Lion, Konstantin and Lipsunen, Werner and Lischner, Johannes and Litman, Yair and Liu, Chi and Liu, Qing-Long and Logsdail, Andrew J. and Lorke, Michael and Lou, Zekun and Mandzhieva, Iuliia and Marek, Andreas and Margraf, Johannes T. and Maurer, Reinhard J. and Melson, Tobias and Merz, Florian and Meyer, J\"{o}rg and Michelitsch, Georg S. and Mizoguchi, Teruyasu and Moerman, Evgeny and Morgan, Dylan and Morgenstein, Jack and Moussa, Jonathan and Nair, Akhil S. and Nemec, Lydia and Oberhofer, Harald and Otero-de-la-Roza, Alberto and Panad\'{e}s-Barrueta, Ram\'{o}n L. and Patlolla, Thanush and Pogodaeva, Mariia and P\"{o}ppl, Alexander and Price, Alastair J. A. and Purcell, Thomas A. R. and Quan, Jingkai and Raimbault, Nathaniel and Rampp, Markus and Rasim, Karsten and Redmer, Ronald and Ren, Xinguo and Reuter, Karsten and Richter, Norina A. and Ringe, Stefan and Rinke, Patrick and Rittmeyer, Simon P. and Rivera-Arrieta, Herzain I. and Ropo, Matti and Rossi, Mariana and Ruiz, Victor and Rybin, Nikita and Sanfilippo, Andrea and Scheffler, Matthias and Scheurer, Christoph and Schober, Christoph and Schubert, Franziska and Shen, Tonghao and Shepard, Christopher and Shang, Honghui and Shibata, Kiyou and Sobolev, Andrei and Song, Ruyi and Soon, Aloysius and Speckhard, Daniel T. and Stishenko, Pavel V. and Tahir, Muhammad and Takahara, Izumi and Tang, Jun and Tang, Zechen and Theis, Thomas and Theiss, Franziska and Tkatchenko, Alexandre and Todorovi\'{c}, Milica and Trenins, George and Unke, Oliver T. and V\'{a}zquez-Mayagoitia, \'{A}lvaro and van Vuren, Oscar and Waldschmidt, Daniel and Wang, Han and Wang, Yanyong and Wieferink, J\"{u}rgen and Wilhelm, Jan and Woodley, Scott and Xu, Jianhang and Xu, Yong and Yao, Yi and Yao, Yingyu and Yoon, Mina and Yu, Victor Wen-zhe and Yuan, Zhenkun and Zacharias, Marios and Zhang, Igor Ying and Zhang, Min-Ye and Zhang, Wentao and Zhao, Rundong and Zhao, Shuo and Zhou, Ruiyi and Zhou, Yuanyuan and Zhu, Tong},
  title = {Roadmap on Advancements of the {FHI}-aims Software Package},
  note = {arXiv:2505.00125},
  year = {2025}
}

@article{MaterialsCloud,
  title = {Materials Cloud,  a platform for open computational science},
  volume = {7},
  ISSN = {2052-4463},
  url = {http://dx.doi.org/10.1038/s41597-020-00637-5},
  DOI = {10.1038/s41597-020-00637-5},
  number = {1},
  journal = {Scientific Data},
  publisher = {Springer Science and Business Media LLC},
  author = {Talirz,  Leopold and Kumbhar,  Snehal and Passaro,  Elsa and Yakutovich,  Aliaksandr V. and Granata,  Valeria and Gargiulo,  Fernando and Borelli,  Marco and Uhrin,  Martin and Huber,  Sebastiaan P. and Zoupanos,  Spyros and Adorf,  Carl S. and Andersen,  Casper Welzel and Sch\"{u}tt,  Ole and Pignedoli,  Carlo A. and Passerone,  Daniele and VandeVondele,  Joost and Schulthess,  Thomas C. and Smit,  Berend and Pizzi,  Giovanni and Marzari,  Nicola},
  year = {2020},
  month = sep 
}

@article{r2scan,
  title = {Accurate and Numerically Efficient r2SCAN Meta-Generalized Gradient Approximation},
  volume = {11},
  ISSN = {1948-7185},
  url = {http://dx.doi.org/10.1021/acs.jpclett.0c02405},
  DOI = {10.1021/acs.jpclett.0c02405},
  number = {19},
  journal = {The Journal of Physical Chemistry Letters},
  publisher = {American Chemical Society (ACS)},
  author = {Furness,  James W. and Kaplan,  Aaron D. and Ning,  Jinliang and Perdew,  John P. and Sun,  Jianwei},
  year = {2020},
  month = sep,
  pages = {8208–8215}
}

@article{mad-dataset,
  title = {Massive Atomic Diversity: a compact universal dataset for atomistic machine learning},
  volume = {12},
  ISSN = {2052-4463},
  url = {http://dx.doi.org/10.1038/s41597-025-06109-y},
  DOI = {10.1038/s41597-025-06109-y},
  number = {1},
  journal = {Scientific Data},
  publisher = {Springer Science and Business Media LLC},
  author = {Mazitov,  Arslan and Chorna,  Sofiia and Fraux,  Guillaume and Bercx,  Marnik and Pizzi,  Giovanni and De,  Sandip and Ceriotti,  Michele},
  year = {2025},
  month = nov 
}

@article{pozdnyakov2023smooth,
  title={Smooth, exact rotational symmetrization for deep learning on point clouds},
  author={Pozdnyakov, Sergey and Ceriotti, Michele},
  journal={Advances in Neural Information Processing Systems},
  volume={36},
  pages={79469--79501},
  year={2023}
}

@article{bigi2026pushing,
  doi = {10.1088/2632-2153/ae6417},
  url = {https://doi.org/10.1088/2632-2153/ae6417},
  year = {2026},
  month = {jun},
  publisher = {IOP Publishing},
  volume = {7},
  number = {3},
  pages = {035051},
  author = {Bigi, Filippo and Pegolo, Paolo and Mazitov, Arslan and Schmidt, Jonathan and Ceriotti, Michele},
  title = {Pushing the limits of unconstrained machine-learned interatomic potentials},
  journal = {Machine Learning: Science and Technology},
}

@article{mazitov2025pet,
  title={PET-MAD as a lightweight universal interatomic potential for advanced materials modeling},
  author={Mazitov, Arslan and Bigi, Filippo and Kellner, Matthias and Pegolo, Paolo and Tisi, Davide and Fraux, Guillaume and Pozdnyakov, Sergey and Loche, Philip and Ceriotti, Michele},
  journal={Nature Communications},
  volume={16},
  number={1},
  pages={10653},
  year={2025},
  publisher={Nature Publishing Group UK London}
}

@article{barroso2024open,
  title={Open materials 2024 (omat24) inorganic materials dataset and models},
  author={Barroso-Luque, Luis and Shuaibi, Muhammed and Fu, Xiang and Wood, Brandon M and Dzamba, Misko and Gao, Meng and Rizvi, Ammar and Zitnick, C Lawrence and Ulissi, Zachary W},
  journal={arXiv preprint arXiv:2410.12771},
  year={2024}
}

@article{adamw,
  title={Decoupled weight decay regularization},
  author={Loshchilov, Ilya and Hutter, Frank},
  journal={arXiv preprint arXiv:1711.05101},
  year={2017}
}

@article{cosineannealing,
  title={Sgdr: Stochastic gradient descent with warm restarts},
  author={Loshchilov, Ilya and Hutter, Frank},
  journal={arXiv preprint arXiv:1608.03983},
  year={2016}
}

@article{bigi2025metatensor,
  title={Metatensor and metatomic: foundational libraries for interoperable atomistic machine learning},
  author={Bigi, Filippo and Abbott, Joseph W and Loche, Philip and Mazitov, Arslan and Tisi, Davide and Langer, Marcel F and Goscinski, Alexander and Pegolo, Paolo and Chong, Sanggyu and Goswami, Rohit and others},
  journal={arXiv preprint arXiv:2508.15704},
  year={2025}
}

@article{matpes,
  title={A foundational potential energy surface dataset for materials},
  author={Kaplan, Aaron D and Liu, Runze and Qi, Ji and Ko, Tsz Wai and Deng, Bowen and Riebesell, Janosh and Ceder, Gerbrand and Persson, Kristin A and Ong, Shyue Ping},
  journal={arXiv preprint arXiv:2503.04070},
  year={2025}
}

@article{bigi2024prediction,
  title={A prediction rigidity formalism for low-cost uncertainties in trained neural networks},
  author={Bigi, Filippo and Chong, Sanggyu and Ceriotti, Michele and Grasselli, Federico},
  journal={Machine Learning: Science and Technology},
  volume={5},
  number={4},
  pages={045018},
  year={2024},
  publisher={IOP Publishing}
}

@article{kellner2024uncertainty,
  title={Uncertainty quantification by direct propagation of shallow ensembles},
  author={Kellner, Matthias and Ceriotti, Michele},
  journal={Machine Learning: Science and Technology},
  volume={5},
  number={3},
  pages={035006},
  year={2024},
  publisher={IOP Publishing}
}

@article{ase,
  title={The atomic simulation environment—a Python library for working with atoms},
  author={Hjorth Larsen, Ask and J{\o}rgen Mortensen, Jens and Blomqvist, Jakob and Castelli, Ivano E and Christensen, Rune and Du{\l}ak, Marcin and Friis, Jesper and Groves, Michael N and Hammer, Bj{\o}rk and Hargus, Cory and others},
  journal={Journal of Physics: Condensed Matter},
  volume={29},
  number={27},
  pages={273002},
  year={2017},
  publisher={IOP Publishing}
}

@article{lammps,
  title={LAMMPS-a flexible simulation tool for particle-based materials modeling at the atomic, meso, and continuum scales},
  author={Thompson, Aidan P and Aktulga, H Metin and Berger, Richard and Bolintineanu, Dan S and Brown, W Michael and Crozier, Paul S and In't Veld, Pieter J and Kohlmeyer, Axel and Moore, Stan G and Nguyen, Trung Dac and others},
  journal={Computer physics communications},
  volume={271},
  pages={108171},
  year={2022},
  publisher={Elsevier}
}

@article{kokkos,
  author={Trott, Christian R. and Lebrun-Grandié, Damien and Arndt, Daniel and Ciesko, Jan and Dang, Vinh and Ellingwood, Nathan and Gayatri, Rahulkumar and Harvey, Evan and Hollman, Daisy S. and Ibanez, Dan and Liber, Nevin and Madsen, Jonathan and Miles, Jeff and Poliakoff, David and Powell, Amy and Rajamanickam, Sivasankaran and Simberg, Mikael and Sunderland, Dan and Turcksin, Bruno and Wilke, Jeremiah},
  journal={IEEE Transactions on Parallel and Distributed Systems},
  title={Kokkos 3: Programming Model Extensions for the Exascale Era},
  year={2022},
  volume={33},
  number={4},
  pages={805-817},
  doi={10.1109/TPDS.2021.3097283}}

@article{bryenton2023delocalization,
  title={Delocalization error: The greatest outstanding challenge in density-functional theory},
  author={Bryenton, Kyle R and Adeleke, Adebayo A and Dale, Stephen G and Johnson, Erin R},
  journal={Wiley Interdisciplinary Reviews: Computational Molecular Science},
  volume={13},
  number={2},
  pages={e1631},
  year={2023},
  publisher={Wiley Online Library}
}

@article{Perdew1996,
  title = {Generalized Gradient Approximation Made Simple},
  volume = {77},
  ISSN = {1079-7114},
  url = {http://dx.doi.org/10.1103/PhysRevLett.77.3865},
  DOI = {10.1103/physrevlett.77.3865},
  number = {18},
  journal = {Physical Review Letters},
  publisher = {American Physical Society (APS)},
  author = {Perdew,  John P. and Burke,  Kieron and Ernzerhof,  Matthias},
  year = {1996},
  month = oct,
  pages = {3865–3868}
}

@article{Sun2015,
  title = {Strongly Constrained and Appropriately Normed Semilocal Density Functional},
  volume = {115},
  ISSN = {1079-7114},
  url = {http://dx.doi.org/10.1103/PhysRevLett.115.036402},
  DOI = {10.1103/physrevlett.115.036402},
  number = {3},
  journal = {Physical Review Letters},
  publisher = {American Physical Society (APS)},
  author = {Sun,  Jianwei and Ruzsinszky,  Adrienn and Perdew,  John P.},
  year = {2015},
  month = jul 
}

@article{chong2023robustness,
  title={Robustness of local predictions in atomistic machine learning models},
  author={Chong, Sanggyu and Grasselli, Federico and Ben Mahmoud, Chiheb and Morrow, Joe D and Deringer, Volker L and Ceriotti, Michele},
  journal={Journal of Chemical Theory and Computation},
  volume={19},
  number={22},
  pages={8020--8031},
  year={2023},
  publisher={ACS Publications}
}

@article{chong2025prediction,
  title={Prediction rigidities for data-driven chemistry},
  author={Chong, Sanggyu and Bigi, Filippo and Grasselli, Federico and Loche, Philip and Kellner, Matthias and Ceriotti, Michele},
  journal={Faraday Discussions},
  volume={256},
  pages={322--344},
  year={2025},
  publisher={Royal Society of Chemistry}
}

@article{alexandria,
  title={Improving machine-learning models in materials science through large datasets},
  author={Schmidt, Jonathan and Cerqueira, Tiago FT and Romero, Aldo H and Loew, Antoine and J{\"a}ger, Fabian and Wang, Hai-Chen and Botti, Silvana and Marques, Miguel AL},
  journal={Materials Today Physics},
  volume={48},
  pages={101560},
  year={2024},
  publisher={Elsevier}
}

@article{omat,
  title={Open materials 2024 (omat24) inorganic materials dataset and models},
  author={Barroso-Luque, Luis and Shuaibi, Muhammed and Fu, Xiang and Wood, Brandon M and Dzamba, Misko and Gao, Meng and Rizvi, Ammar and Zitnick, C Lawrence and Ulissi, Zachary W},
  journal={arXiv preprint arXiv:2410.12771},
  year={2024}
}

@article{omol,
  title={The open molecules 2025 (omol25) dataset, evaluations, and models},
  author={Levine, Daniel S and Shuaibi, Muhammed and Spotte-Smith, Evan Walter Clark and Taylor, Michael G and Hasyim, Muhammad R and Michel, Kyle and Batatia, Ilyes and Cs{\'a}nyi, G{\'a}bor and Dzamba, Misko and Eastman, Peter and others},
  journal={arXiv preprint arXiv:2505.08762},
  year={2025}
}

@article{wbm,
  title={Predicting stable crystalline compounds using chemical similarity},
  author={Wang, Hai-Chen and Botti, Silvana and Marques, Miguel AL},
  journal={npj Computational Materials},
  volume={7},
  number={1},
  pages={12},
  year={2021},
  publisher={Nature Publishing Group UK London}
}

@article{Hart2008,
  title = {Algorithm for generating derivative structures},
  volume = {77},
  ISSN = {1550-235X},
  url = {http://dx.doi.org/10.1103/PhysRevB.77.224115},
  DOI = {10.1103/physrevb.77.224115},
  number = {22},
  journal = {Physical Review B},
  publisher = {American Physical Society (APS)},
  author = {Hart,  Gus L. W. and Forcade,  Rodney W.},
  year = {2008},
  month = jun 
}

@article{Kolli2020,
  title = {Discovering hierarchies among intermetallic crystal structures},
  author = {Kolli, Sanjeev Krishna and Natarajan, Anirudh Raju and Thomas, John C. and Pollock, Tresa M. and Van der Ven, Anton},
  journal = {Phys. Rev. Mater.},
  volume = {4},
  issue = {11},
  pages = {113604},
  numpages = {12},
  year = {2020},
  month = {Nov},
  publisher = {American Physical Society},
  doi = {10.1103/PhysRevMaterials.4.113604},
  url = {https://link.aps.org/doi/10.1103/PhysRevMaterials.4.113604}
}

@article{kotha+acs2023,
author = {Kothakonda, Manish and Kaplan, Aaron D. and Isaacs, Eric B. and Bartel, Christopher J. and Furness, James W. and Ning, Jinliang and Wolverton, Chris and Perdew, John P. and Sun, Jianwei},
title = {Testing the r2SCAN Density Functional for the Thermodynamic Stability of Solids with and without a van der Waals Correction},
journal = {ACS Materials Au},
volume = {3},
number = {2},
pages = {102-111},
year = {2023},
doi = {10.1021/acsmaterialsau.2c00059},

URL = { 
    
        https://doi.org/10.1021/acsmaterialsau.2c00059
    
    

},
eprint = { 
    
        https://doi.org/10.1021/acsmaterialsau.2c00059
    
    

}

}

@article{sun+nchem2016, 
year = {2016}, 
title = {{Accurate first-principles structures and energies of diversely bonded systems from an efficient density functional}}, 
author = {Sun, Jianwei and Remsing, Richard C. and Zhang, Yubo and Sun, Zhaoru and Ruzsinszky, Adrienn and Peng, Haowei and Yang, Zenghui and Paul, Arpita and Waghmare, Umesh and Wu, Xifan and Klein, Michael L. and Perdew, John P.}, 
journal = {Nature Chemistry}, 
issn = {1755-4330}, 
doi = {10.1038/nchem.2535}, 
pmid = {27554409}, 
pages = {831--836}, 
number = {9}, 
volume = {8}, 
}

@article{Liu2024,
  title = {Assessing r2SCAN meta-GGA functional for structural parameters,  cohesive energy,  mechanical modulus,  and thermophysical properties of 3d,  4d,  and 5d transition metals},
  volume = {160},
  ISSN = {1089-7690},
  url = {http://dx.doi.org/10.1063/5.0176415},
  DOI = {10.1063/5.0176415},
  number = {2},
  journal = {The Journal of Chemical Physics},
  publisher = {AIP Publishing},
  author = {Liu,  Haoliang and Bai,  Xue and Ning,  Jinliang and Hou,  Yuxuan and Song,  Zifeng and Ramasamy,  Akilan and Zhang,  Ruiqi and Li,  Yefei and Sun,  Jianwei and Xiao,  Bing},
  year = {2024},
  month = jan 
}

@article{Thiemann2024,
  title = {Introduction to machine learning potentials for atomistic simulations},
  volume = {37},
  ISSN = {1361-648X},
  url = {http://dx.doi.org/10.1088/1361-648X/ad9657},
  DOI = {10.1088/1361-648x/ad9657},
  number = {7},
  journal = {Journal of Physics: Condensed Matter},
  publisher = {IOP Publishing},
  author = {Thiemann,  Fabian L and O’Neill,  Niamh and Kapil,  Venkat and Michaelides,  Angelos and Schran,  Christoph},
  year = {2024},
  month = dec,
  pages = {073002}
}

@article{Behler2016,
  title = {Perspective: Machine learning potentials for atomistic simulations},
  volume = {145},
  ISSN = {1089-7690},
  url = {http://dx.doi.org/10.1063/1.4966192},
  DOI = {10.1063/1.4966192},
  number = {17},
  journal = {The Journal of Chemical Physics},
  publisher = {AIP Publishing},
  author = {Behler,  J\"{o}rg},
  year = {2016},
  month = nov 
}

@article{Jacobs2025,
  title = {A practical guide to machine learning interatomic potentials – Status and future},
  volume = {35},
  ISSN = {1359-0286},
  url = {http://dx.doi.org/10.1016/j.cossms.2025.101214},
  DOI = {10.1016/j.cossms.2025.101214},
  journal = {Current Opinion in Solid State and Materials Science},
  publisher = {Elsevier BV},
  author = {Jacobs,  Ryan and Morgan,  Dane and Attarian,  Siamak and Meng,  Jun and Shen,  Chen and Wu,  Zhenghao and Xie,  Clare Yijia and Yang,  Julia H. and Artrith,  Nongnuch and Blaiszik,  Ben and Ceder,  Gerbrand and Choudhary,  Kamal and Csanyi,  Gabor and Cubuk,  Ekin Dogus and Deng,  Bowen and Drautz,  Ralf and Fu,  Xiang and Godwin,  Jonathan and Honavar,  Vasant and Isayev,  Olexandr and Johansson,  Anders and Kozinsky,  Boris and Martiniani,  Stefano and Ong,  Shyue Ping and Poltavsky,  Igor and Schmidt,  KJ and Takamoto,  So and Thompson,  Aidan P. and Westermayr,  Julia and Wood,  Brandon M.},
  year = {2025},
  month = mar,
  pages = {101214}
}

@article{Chen2017,
  title = {Ab initio theory and modeling of water},
  volume = {114},
  ISSN = {1091-6490},
  url = {http://dx.doi.org/10.1073/pnas.1712499114},
  DOI = {10.1073/pnas.1712499114},
  number = {41},
  journal = {Proceedings of the National Academy of Sciences},
  publisher = {Proceedings of the National Academy of Sciences},
  author = {Chen,  Mohan and Ko,  Hsin-Yu and Remsing,  Richard C. and Calegari Andrade,  Marcos F. and Santra,  Biswajit and Sun,  Zhaoru and Selloni,  Annabella and Car,  Roberto and Klein,  Michael L. and Perdew,  John P. and Wu,  Xifan},
  year = {2017},
  month = sep,
  pages = {10846–10851}
}

\setcounter{figure}{0}
\renewcommand{\thefigure}{S\arabic{figure}}
\onecolumngrid\clearpage

\section*{SUPPLEMENTARY INFORMATION}

\subsection{LLPR dataset cleaning}

\begin{figure}[tbh]
    \centering
    \includegraphics[width=0.8\columnwidth]{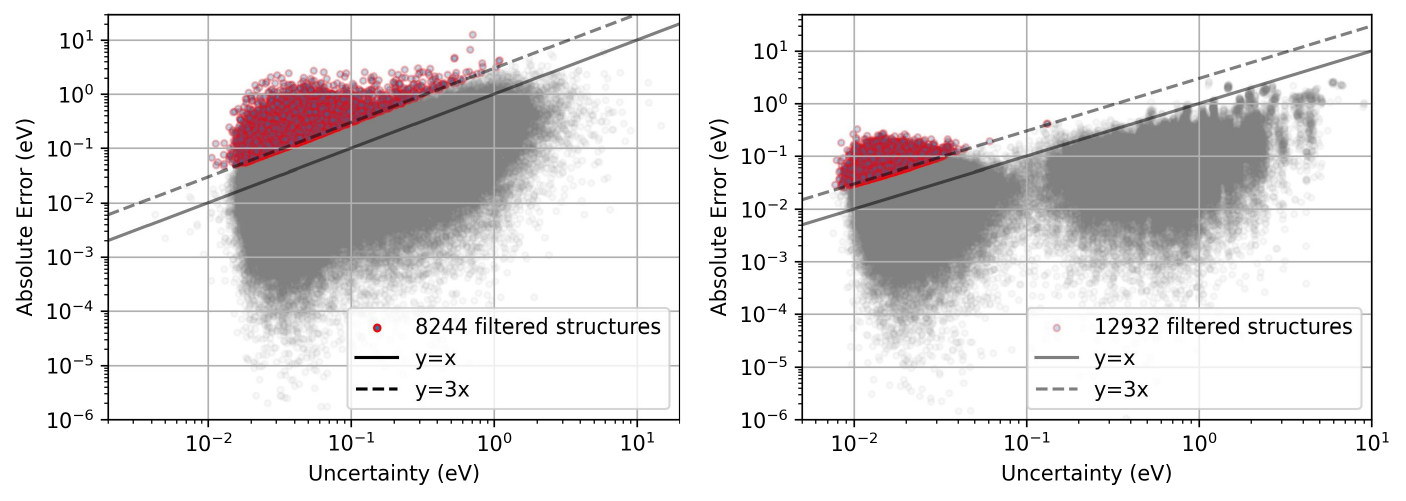}
    \caption{Visualization of the dataset cleaning based on predicted LLPR uncertainties. Left and right panels represent two independent iterations of cleaning done for the MAD-1.5 dataset, and for the new subsets in the MAD-1.6 dataset (\emph{trimers-extended}, and \emph{MAD-cat}). Each iteration involved a separate LLPR model training and calibration on aforementioned data. All the structures for which the actual absolute error in energy predictions is 3 times higher than the predicted energy uncertainty are filtered out. A total of 21 176 structures filtered from the pre-cleaned version of the dataset are marked with red circles. The rest of the dataset is represented with gray dots.}
    \label{fig:llpr-cleaning}
\end{figure}

\subsection{Basis set details in electronic-structure calculations}

\begin{figure}[tb]
    \centering    \includegraphics[width=0.8\linewidth]{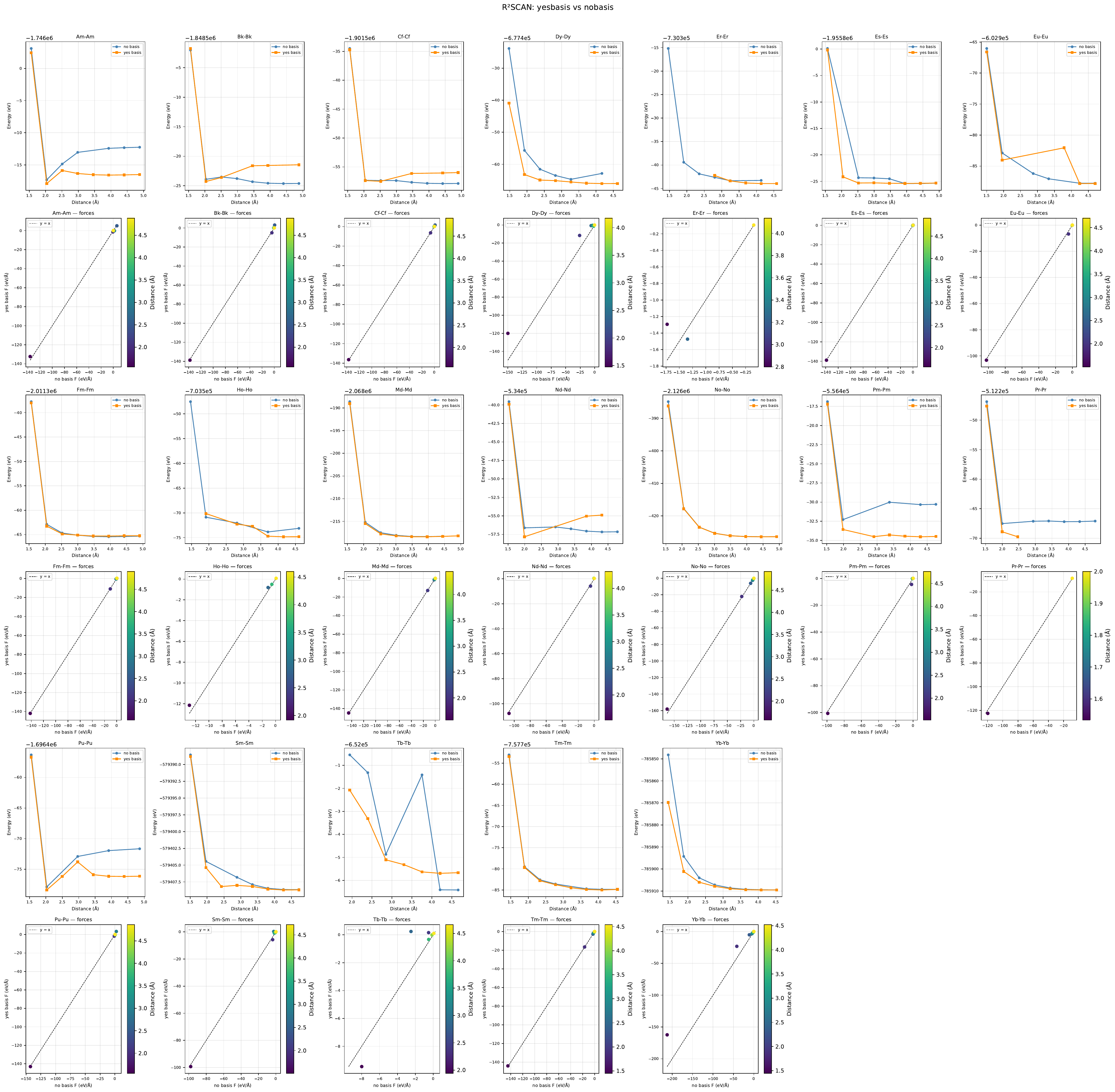}
    \caption{Each panel shows the potential energy curve (top) and a parity plot of interaction forces (bottom) for a homonuclear dimer of the indicated lanthanide or actinide element, computed with (orange) and without (blue) the confined 5d/6d function that appears as an addition to the minimal basis in the FHI-aims default species files. The colorbar in the paritiy plots indicates interatomic distance.}
    \label{fig:dimer_comparison}
\end{figure}
For a subset of lanthanide and actinide elements (Pr–Yb, Pu–No, excluding Ce), the default NAO basis sets include a confined 5d/6d function as a necessary addition to the minimal basis, reflecting the near-degeneracy of the unoccupied d manifold with the partially filled 4f/5f states. However, in non-spin-polarized calculations of multi-component solids, these functions were found to make SCF convergence more difficult; we therefore omitted them. To assess the physical impact of this choice, we computed homonuclear dimer potential energy curves for all affected elements with and without the confined d function (Fig. \ref{fig:dimer_comparison}). The RMSE in forces varies across elements, from ~0.1 eV/Å (Es, Er) to ~19 eV/Å (Yb) and ~14 eV/Å (Dy), indicating that the basis removal has a non-uniform impact. We note that these differences are evaluated on homonuclear dimers, which represent a sensitive test case; the impact on the multi-component solid-state systems in this database may differ. Nevertheless, users should be aware that results for these elements carry additional basis-related uncertainty. The removal of the confined function is therefore to be intended as a pragmatic choice motivated by convergence robustness  rather than physical optimality, and we recommend users exercise caution when employing this dataset for applications where the d-manifold contribution is expected to be significant.

\end{document}